\documentclass[journal=jctcce,manuscript=article]{achemso}
\pdfoutput=1
\usepackage{graphicx}%to includegraphics 
\usepackage{hyperref}
\usepackage{txfonts}
\usepackage{color}
\usepackage{xcolor}
\usepackage{amstext}
\hypersetup{
    colorlinks,
    linkcolor={red!50!black},
    citecolor={blue!50!black},
    urlcolor={blue!80!black}
}

\title{Exactly-embedded multiconfigurational self-consistent field theory using density matrix embedding: the localized active space self-consistent field method}

\date{\today}
\author{Matthew R.\ Hermes}
\email{herme068@umn.edu}
\author{Laura Gagliardi}
\email{gagliard@umn.edu}
\affiliation{Department of Chemistry, Chemical Theory Center, and The Minnesota Supercomputing Institute, University of Minnesota, Minneapolis, MN, 55455}

\begin{document}

\begin{abstract}
Density matrix embedding theory (DMET) is a fully quantum-mechanical embedding method which shows great promise as a method of defeating the inherent exponential cost
scaling of multiconfigurational wave function-based calculations by breaking large systems into smaller, coupled subsystems.
However, we recently [\emph{JCTC} \textbf{2018}, \emph{14}, 1960] encountered evidence that the approximate single-determinantal bath picture inherent to DMET
is sometimes problematic when the complete active space self-consistent field (CASSCF) is used as a solver and the method is applied to realistic models of strongly-correlated molecules.
Here, we show this problem can be defeated by generalizing DMET to use a multiconfigurational wave function as a bath
\emph{without} sacrificing practically attractive features of DMET, such as a second-quantization form of the embedded subsystem Hamiltonian,
by dividing the active space into unentangled active subspaces each localized to one fragment. We introduce the term localized active space (LAS) to refer to this kind of wave function.
The LAS bath wave function can be obtained by the DMET algorithm itself in a self-consistent manner, and we refer to this approach, introduced here for the first time, as the localized
active space self-consistent field (LASSCF) method.
LASSCF exploits a modified DMET algorithm, but it requires no ambiguous error function minimization, produces a whole-molecule wave function with exact embedding, is variational, and it reproduces
full-molecule CASSCF in cases where comparable DMET calculations fail. 
Our results for test calculations on the nitrogen double-bond dissociation potential energy curves of several diazene molecules
suggest that LASSCF can be an appropriate starting point for a perturbative treatment.
Outside of the context of embedding, the LAS wave function is inherently an attractive alternative to a CAS wave function
because of its favorable cost scaling, which is exponential only with respect to the size of individual fragment active subspaces, rather than the whole
active space of the entire system.
\end{abstract}
\maketitle

\section{Introduction\label{intro}}

Embedding models\cite{Jacob2014,Wesolowski2015,Collins2015,Raghavachari2015,Sun2016} are designed to facilitate the application of accurate
quantum-chemistry approximations to chemical systems
otherwise too large for the requisite calculations to be practical. They separate the physics of a chemical system into those characterizing
one or several ``small'' center(s) of interest,
variously referred to as subsystems, fragments, or impurities,
and ``large'' environments, where the latter are considered less sensitive to approximation.
The environments can be modeled as a collection of electrostatic point
charges,\cite{Li2007,Isegawa2013,Collins2014,Kamiya2008,Richard2012} as a more general
external potential obtained by differentiation of density functionals,\cite{Cortona1991,Wesolowskv1993}
as a frequency-dependent self-energy term,\cite{Georges1996,Kananenka2015}
or as a small number of linear combinations
of fully quantum-mechanical environment states (``bath states'').
This last form characterizes density matrix embedding theory (DMET).\cite{Knizia2012,Wouters2016,Wouters2017}

The concept of embedding resembles the concept of the ``active space,''\cite{ROOS2005725,Schmidt1998} which is ubiquitious in quantum-chemistry models of strong electron correlation,
except that active spaces are defined in terms of energy and interaction strength between orbitals, rather than in terms of real-space length scales.
In active-space methods such as complete,\cite{Roos1980} restricted,\cite{Olsen1988,Malmqvist1990} generalized,\cite{Ma2011} 
or occupation-restricted multiple\cite{Ivanic2003} active-space self-consistent field (CASSCF, RASSCF, GASSCF, ORMAS-SCF),
the ``fragment'' is the collection of active molecular orbitals (MOs). A large configuration interaction (CI) calculation is carried out in the Fock space of the fragment,
while the ``environment'' is modeled as a single-determinantal wave function in the Fock space of non-active orbitals.
This approach is known as multi-configurational self-consistent field (MC-SCF).
Note that in MC-SCF, unlike most real-space embedding methods, the actual partition of a system into ``fragment'' and ``environment'' is itself variationally optimized.

Active-space methods such as CASSCF suffer from exponential computational cost scaling in the size of the active space, and are not usually
practical methods for calculations of large molecules or materials such as metal-organic frameworks.\cite{Horike2009,Lee2009a,Odoh2015,Coudert2016,Yan2017,Bernales2018} When the effects of strong electron
correlation are investigated for these systems, the usual procedure is to perform active-space calculations on small model molecules or clusters.\cite{Odoh2015}
Practical MC-SCF calculations on large systems could be greatly facilitated by the combination of the active-space approach with real-space embedding via methods
such as DMET. 

DMET is based on the concept of a Schmidt decomposition,\cite{Schmidt1907,Peschel2012} which exactly expresses the wave function of a large system in terms
of a state space no larger than twice the size of a state space of any given fragment, by performing singular-value decomposition (SVD)
of the wave function's coefficient tensor.
This generates an ``impurity'' state space, consisting of fragment states and bath states, which is guaranteed
to contain the trial wave function used to generate it. Projecting the molecular Hamiltonian into the impurity subspace makes the solution of the Schr\"{o}dinger equation
vastly more computationally affordable. Additionally, this embedding method, unlike electrostatic embedding or
QM/MM techniques,\cite{Waecziel1976,Lin2007}
explicitly accounts for the effects of entanglement and the whole model remains quantum-mechanical.

Standard DMET uses a single-determinantal trial wave function to generate the impurity space,
in which it then obtains a correlated, multi-determinantal wave function using, for example, full configuration interaction (FCI) methods.
Even though single determinants, by definition, cannot model strongly-correlated systems with even qualitative accuracy,
DMET calculations have repeatedly been demonstrated to produce accurate results for
stretched hydrogen chains, sheets, and rings,\cite{Knizia2013,Wouters2016,Pham2018} as well as strongly-correlated Hubbard models\cite{Knizia2012}
and other abstract strongly-correlated model systems.\cite{Sandhoefer2016} Clearly, the reliance on a single-determinantal trial wave function
is not necessarily fatal for DMET. However, note that the above examples exhibit high symmetry and are fairly abstract in comparison to 
real molecules and solids, and the literature currently has fewer records of DMET being applied to more sophisticated molecules with large basis sets.

We recently investigated\cite{Pham2018} the performance of DMET using CASSCF as an ``impurity solver'' for
the projected Schr\"{o}dinger equations. One goal was to determine if this ``CAS-DMET'' approach could generate CASSCF-quality
descriptions of molecular models of strong correlation; specifically, homolytically broken nitrogen-nitrogen double bonds. 
The quality of the CAS-DMET results depended on the system and on the level of approximation in DMET, but
for the case of the dissociation of the nitrogen-nitrogen double bond of the azomethane molecule, a \emph{post-hoc} modification
of DMET, in which all but a small number of the most strongly-entangled bath orbitals of the nitrogen fragment were dropped from the impurity subspace,
 was necessary in order to generate a qualitatively accurate potential energy curve. This hints that the high accuracy of DMET calculations performed on 
simple, high-symmetry models of strong electron correlation may not necessarily generalize robustly to more detailed chemical models,
to impurity solvers less complete than FCI, or to both.

Could the single-determinantal trial wave function of standard DMET be replaced with a correlated wave function, and would this render results
for strongly-correlated molecules more accurate?
Although correlated trial wave functions have occasionally been used in the literature,\cite{Tsuchimochi2015,Fan2015,Gunst2017} this approach carries a drawback.
The single-determinantal trial wave function of DMET guarantees that the impurity subspace generated by Schmidt decomposition is a simple Fock space
of orbitals occupied by an integer number of electrons. This is a significant practical appeal of DMET, since \emph{ab initio} quantum chemistry
methods such as CASSCF are defined in these terms. Thus, DMET can be interfaced with existing performant implementations of powerful quantum chemistry
approximations with very little modification. But if a correlated trial wave function is used, then the impurity space is not necessarily a Fock space
and it does not even necessarily contain an integer number of electrons.

However, a single determinant is not the only form of a trial wave function which generates an impurity subspace corresponding
to a Fock space of orbitals. In this work, we define for the first time a localized active space (LAS) wave function as the lowest on a systematic hierarchy
of MC-SCF wave functions, wherein the active space is split into unentangled subspaces located on different fragments. 
We will show that, so long as the choice of fragments obeys a certain constraint, the numerical recipe for Schmidt decomposition of a single determinant 
and for projection of the molecular Hamiltonian into the impurity space also applies to a LAS trial wave function \emph{without modification}.
Thus, the standard DMET algorithm can be adapted straightforwardly to the variational optimization of a LAS wave function,
provided the orbitals defining the fragments are allowed to ``relax,'' as described below.

We define this new approach as localized active-space self-consistent field (LASSCF). LASSCF is a union of DMET and MC-SCF which
breaks the relaxation of active orbital coefficients and CI vectors into coupled localized subsystem problems. As the name implies,
LASSCF generates a true wave function and the embedding is exact at convergence, unlike standard DMET, which must make use of a one-body correlation potential
and an arbitrary error function to optimize its trial wave function. We test LASSCF and CAS-DMET on several model molecules,
including azomethane, to further examine whether CAS-DMET may fail for strongly-correlated systems and under what conditions, and whether LASSCF
constitutes a qualitative improvement. We find that LASSCF provides results extremely close to whole-molecule CASSCF for systems for which standard DMET
barely functions - the bath states generated by Schmidt decomposition of the restricted Hartree--Fock (RHF) wave function are qualitatively bad,
and the minimization of the error function appears to be impossible.

The remainder of this work is organized as follows:\ Sec.\ \ref{dmet_background} reviews standard DMET, and Sec.\ \ref{schmidt_analysis} analyzes the Schmidt decomposition
to determine how LASSCF can be efficiently implemented. Sec.\ \ref{lasscf} then provides essential technical details of the LASSCF algorithm. Sec.\ \ref{results_discussion} presents
test calculations of LASSCF as well as standard DMET on three model molecules, and analyzes the comparative performance.
Sec.\ \ref{conclusions} offers some conclusions. An appendix discusses technical details of our LASSCF implementation.

\section{Theory}

\subsection{DMET review\label{dmet_background}}

In DMET, a large molecule or extended system is separated by the user into several
(in the ``democratic''\cite{Wouters2016} formalism)
non-overlapping fragments, here indexed with capital Roman letters $K,L,M,N,\ldots$.
Each fragment is defined by the partition of the whole system's $M_\mathrm{mol}$ spinorbitals
into $M_{F_K}$ ``fragment'' orbitals, indexed as $f_{Kn}$, $n=1,2,\ldots$, and $M_{E_K}=M_\mathrm{mol}-M_{F_K}$
``environment'' orbitals, $e_{Kn}$. A trial wave function is subjected to a Schmidt decomposition,
\begin{eqnarray}
    |\Psi_\mathrm{tr}\rangle &=& \sum_{\vec{f}_K}^{n_{F_K}} \sum_{\vec{e}_K}^{n_{E_K}} C^{\vec{f}_K}_{\vec{e}_K} |\vec{f}_K\rangle \otimes |\vec{e}_K\rangle 
    \nonumber \\ 
    &=& \sum_{\vec{f}_K}^{n_{F_K}} \sum_{\vec{e}_K}^{n_{E_K}} \sum_{B_K}^{n_{B_K}}
    U^{\vec{f}_K}_{B_K}\sigma_{B_K}\left(V^{\vec{e}_K}_{B_K}\right)^* |\vec{f}_K\rangle \otimes |\vec{e}_K\rangle
    \nonumber \\
    &=& \sum_{\vec{f}_K}^{n_{F_K}} \sum_{B_K}^{n_{B_K}} U^{\vec{f}_K}_{B_K} \sigma_{B_K} |\vec{f}_K\rangle \otimes |B_K\rangle,
    \label{schmidt_general}
\end{eqnarray}
where $\vec{f}_K=\{f_{K1},f_{K2},\ldots\}$ and $\vec{e}_K=\{e_{K1},e_{K2},\ldots\}=\vec{f}_L\vec{f}_M\cdots$ are lists of occupied fragment and environment orbitals
(i.e., occupancy vectors)
in the determinants $|\vec{f}_K\rangle$ and $|\vec{e}_K\rangle$,
and $n_{B_K}\le \mathrm{min}(n_{F_K},n_{E_K})$ is the number of nonzero singular values ($\sigma_{B_K}$) of the coefficient tensor
$(C^{\vec{f}_K}_{\vec{e}_K})$.
The right-singular vectors of the coefficient tensor ($V^{\vec{e}_K}_{B_K}$)
corresponding to nonzero $\sigma_{B_K}$ generate many-body ``bath states'' $(|B_K\rangle)$;\ each bath state is entangled
to the fragment state generated by the corresponding left-singular vector ($U^{\vec{f}_K}_{B_K}$).
Projecting the molecular Hamiltonian into the fragment-and-bath (``impurity'') subspace,
\begin{eqnarray}
    \hat{H}_{I_K} &\equiv& \hat{P}_{I_K}\hat{H}\hat{P}_{I_K},
    \label{H_I_K}
    \\ 
    \hat{P}_{I_K} &\equiv& \sum_{\vec{f}_K}^{n_{F_K}}\sum_{B_K}^{n_{B_K}} |\vec{f}_K\rangle\otimes|B_K\rangle\langle B_K|\otimes \langle \vec{f}_K|,
    \label{P_I_K}
\end{eqnarray}
results in a reduced-dimensional Schr\"{o}dinger equation that has the trial wave function as one of its solutions,
\begin{eqnarray}
    \hat{H}_{I_K}|\Psi_{I_K}\rangle &=& E_{I_K}|\Psi_{I_K}\rangle = E_\mathrm{tr}|\Psi_\mathrm{tr}\rangle.
    \label{impurity_problem_nomu}
\end{eqnarray}
Equations (\ref{schmidt_general})--(\ref{impurity_problem_nomu}) trace the outline of a self-consistent algorithm for solving the Schr\"{o}dinger equation
of a large system as a coupled set of smaller Schr\"{o}dinger equations, which, at convergence, all have the target eigensolution in common. However, 
without further approximation, this hypothetical algorithm is not practically useful:\ the first step is about as computationally costly as FCI,
since the coefficient tensor grows exponentially with the size of the system.

Standard DMET, instead, utilizes a single-determinantal trial wave function, $|\Phi_\mathrm{tr}\rangle$,
to generate the impurity space [Eq.\ (\ref{schmidt_general})], along with
a correlated post-Hartree--Fock or FCI method to solve the impurity problems [Eq.\ (\ref{impurity_problem_nomu})]. The trial wave function solves
\begin{eqnarray}
    \left(\hat{h} + \sum_K \hat{u}_K\right)|\Phi_\mathrm{tr}\rangle &=& E_\mathrm{tr}|\Phi_\mathrm{tr}\rangle,
    \label{trial_eigenequation}
\end{eqnarray}
where $\hat{h}$ is, e.g., the standard Fock operator from a whole-molecule Hartree--Fock (HF) calculation, and $\hat{u}_K$ is a local correlation
potential affecting the orbitals of the $K$th fragment.
The latter is adjusted iteratively to minimize the difference between the one-body reduced density matrices (1-RDMs) of $|\Phi_\mathrm{tr}\rangle$
and those of the impurity wave functions [$|\Psi_{I_K}\rangle$ from Eq.\ (\ref{impurity_problem_nomu})].
Energies and properties are computed from weighted sums of impurity reduced density matrices,
\begin{eqnarray}
    \gamma^{f_{Kn}}_{f_{Ln}} &=& \frac{1}{2}\left(\{\gamma_{I_K}\}^{f_{Kn}}_{f_{Ln}}+\{\gamma_{I_L}\}^{f_{Kn}}_{f_{Ln}}\right),
    \label{1RDM_output}
    \\
    \gamma^{f_{Kn}f_{Mn}}_{f_{Ln}f_{Nn}} &=& \frac{1}{4}
    \left(\{\gamma_{I_K}\}^{f_{Kn}f_{Mn}}_{f_{Ln}f_{Mn}}+\{\gamma_{I_L}\}^{f_{Kn}f_{Mn}}_{f_{Ln}f_{Nn}}
    \right. \nonumber \\ && \left.
    +\{\gamma_{I_M}\}^{f_{Kn}f_{Mn}}_{f_{Ln}f_{Mn}}+\{\gamma_{I_N}\}^{f_{Kn}f_{Mn}}_{f_{Ln}f_{Nn}}\right),
    \label{2RDM_output} 
\end{eqnarray}
and so forth, where $\gamma_{I_K}$ ($\gamma_{I_L}$, etc.) with $k$ subscript-superscript
columns denotes the $k$-RDM computed from $|\Psi_{I_K}\rangle$ ($|\Psi_{I_L}\rangle$, etc.), respectively.
RDMs computed in this way are not necessarily $N$-representable and standard DMET
does not produce a correlated wave function \emph{per se}. 
Also, because the 1-RDM of a single determinant is constrained to be idempotent and that of a correlated
wave function is not, it is impossible in general to exactly match the 1-RDM of the trial wave function to, i.e., Eq.\ (\ref{1RDM_output}). Instead, a choice of
error function must be made, and unless this error function exactly constrains the trace of the right-hand side of Eq.\ (\ref{1RDM_output}), it becomes necessary to introduce
a global chemical potential [not to be confused with the local
correlation potential of Eq.\ (\ref{trial_eigenequation})] in the solution of the impurity problems,
\begin{eqnarray}
    \left(\hat{H}_{I_K}+\mu\hat{N}_{F_K}\right)|\Psi_{I_K}\rangle &=& E_{I_K}|\Psi_{I_K}\rangle,
    \label{impurity_problem_mu}
\end{eqnarray}
where $\hat{N}_{F_K}$ counts the electrons occupying the $K$th fragment's spinorbitals and $\mu$ is determined so that
$N_\mathrm{mol}$, the total number of electrons in the system, is fixed:
\begin{eqnarray}
    N_\mathrm{mol} &=& \sum_K \langle \Psi_{I_K}|\hat{N}_{F_K}|\Psi_{I_K}\rangle = \sum_K\sum_{f_{Kn}} \{\gamma_{I_K}\}^{f_{Kn}}_{f_{Kn}}.
    \label{chempot_nfix}
\end{eqnarray}
[Equations (\ref{1RDM_output})--(\ref{chempot_nfix}) apply to the ``democratic'' form of DMET, in the terminology of Ref.\ \citenum{Wouters2016};\ alternatively,
in a ``single-embedding'' calculation, only one fragment is defined and its $|\Psi_{I_K}\rangle$ and $E_{I_K}$ are taken as the wave function
and energy of the whole system.]

\begin{figure}[tb]
\includegraphics[scale=1.0]{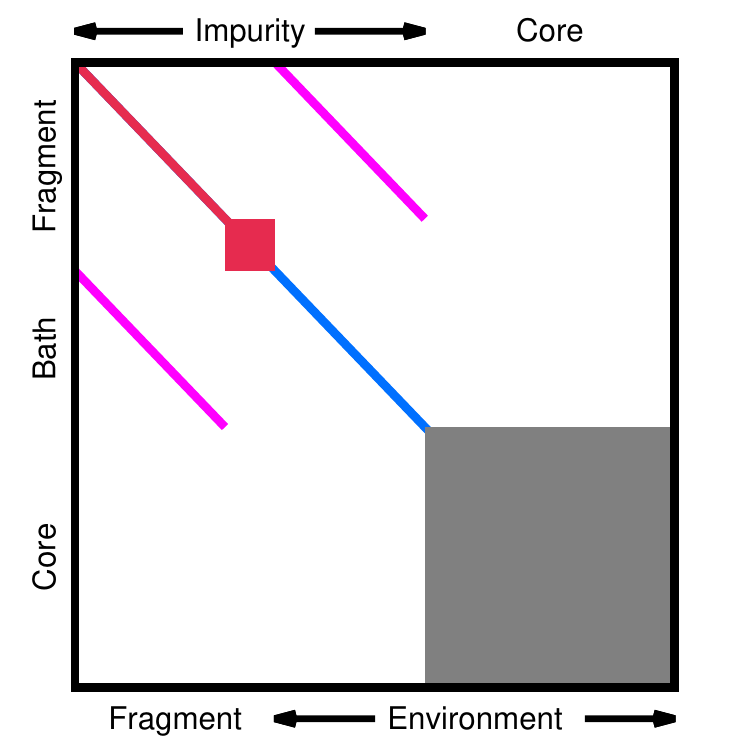}
\caption{Diagrammatic representation of the trial 1-RDM in the embedding
basis generated by the Schmidt decomposition of a single-determinantal trial wave function [Eq.\ (\ref{single_determinant_schmidt})],
generating $M_{B_K}<M_{F_K}$ bath orbitals.
White space represents matrix elements that are identically zero. SVD in Eq.\ (\ref{single_determinant_schmidt})
explicitly diagonalizes the fragment-environment coupling block, as depicted by the purple lines, and indirectly
diagonalizes parts of the fragment-fragment and environment-environment blocks as well (red and blue lines) due to the idempotency
of the 1-RDM. The red and grey squares represent 1-RDM blocks in the left and right null spaces ($\sigma_{b_{K3}}=0$) respectively
of the fragment-environment coupling block. Labels on the borders show the partition of the embedding basis into 3 sets of subspaces;\
the bath with the fragment is the ``impurity,'' and the bath with the core is the ``environment.''
\label{embedding_1rdm_fig}}
\end{figure}

The many-body bath states from Eqs.\ (\ref{schmidt_general}) and (\ref{P_I_K}) ($|B_K\rangle$) for a single determinant, $|\Phi_\mathrm{tr}\rangle$, take the form
\begin{eqnarray}
    |B_K\rangle &\to& |\vec{b}_K\rangle\otimes|\Phi_{C_K}\rangle,
    \label{singdet_mbb}
\end{eqnarray}
where $\vec{b}_K$ lists occupied orbitals in determinant $|\vec{b}_K\rangle$ in the space of $M_{B_K}\le M_{F_K}$ ``bath orbitals,'' indexed as $b_{Kn}$,
and $|\Phi_{C_K}\rangle$ is a single determinant, which all bath states have in common, in the space of $M_{C_K}=M_{E_K}-M_{B_K}$ ``core orbitals,'' indexed as $c_{Kn}$. The
bath and core orbitals are linear combinations of environment orbitals obtained by SVD of the fragment-environment coupling block of the trial wave function's 1-RDM,
\begin{eqnarray}
    \{\gamma_\mathrm{tr}\}^{f_{Kn}}_{e_{Kn}} &=& \sum_{b_{Kn}} u^{f_{Kn}}_{b_{Kn}} \sigma_{b_{Kn}} \left(v^{e_{Kn}}_{b_{Kn}}\right)^*,
    \label{single_determinant_schmidt}
\end{eqnarray}
% Laura: v^eKn_bKn is defined in the first parenthetical below
which has negligible computational cost compared to SVD of the coefficient tensor. 
Transforming the fragment and environment orbitals by the left- ($u^{f_{Kn}}_{b_{Kn}}$) and right-singular vectors ($v^{e_{Kn}}_{b_{Kn}}$) respectively
generates the so-called ``embedding basis;'' the 1-RDM in the embedding basis is depicted in Fig.\ \ref{embedding_1rdm_fig}.
(In practice the transformation of fragment orbitals can be ignored, but we retain this in Fig.\ \ref{embedding_1rdm_fig} in order to depict the diagonalization implied by SVD.)
The bath orbitals arise from right-singular vectors corresponding to nonzero singular value ($\sigma_{b_{Kn}}$) and the core
orbitals span the right null space ($\sigma_{b_{Kn}}=0$).
Just as there can be no more many-body bath states than many-body fragment states ($n_{B_K}\le n_{F_K}$),
there can be no more bath orbitals than fragment orbitals ($M_{B_K}\le M_{F_K}$).
The fragment and bath orbitals combined make $M_{I_K}=M_{F_K}+M_{B_K}=M_\mathrm{mol}-M_C$ ``impurity orbitals,'' indexed as $i_{Kn}$.

The direct-product basis of all $|\vec{b}_K\rangle$ and all $|\vec{f}_K\rangle$ is the complete Fock space
of all impurity orbitals. Thus, the projection of the Hamiltonian into the impurity space
[Eq. (\ref{H_I_K})] is accomplished by representing it in the \emph{single-particle} embedding basis and integrating over the core orbitals,
\begin{eqnarray}
    \hat{H}_{I_K} &=& \tilde{V}_{I_K} + \{\tilde{h}_{I_K}\}^{i_{1}}_{i_{2}}\hat{a}_{i_{1}}^\dagger \hat{a}_{i_{2}}
    + \frac{1}{2}v^{i_{1}i_{3}}_{i_{2}i_{4}}\hat{a}_{i_{1}}^\dagger \hat{a}_{i_{3}}^\dagger \hat{a}_{i_{4}} \hat{a}_{i_{2}},
    \label{H_I}
    \\
    \tilde{V}_{I_K} &=& V_0 + h^{c_{1}}_{c_{2}}\{\gamma_\mathrm{tr}\}^{c_{1}}_{c_{2}}
    + \frac{1}{2} v^{c_{1}c_{3}}_{c_{2}c_{4}}\{\gamma_\mathrm{tr}\}^{c_{1}c_{3}}_{c_{2}c_{4}},
    \label{H0_I}
    \\
    \{\tilde{h}_{I_K}\}^{i_{1}}_{i_{2}} &=& h^{i_{1}}_{i_{2}}
    + \left(v^{i_{1}c_{1}}_{i_{2}c_{2}} - v^{i_{1}c_{1}}_{c_{2}i_{2}}\right)\{\gamma_\mathrm{tr}\}^{c_{1}}_{c_{2}},
    \label{H1_I}
\end{eqnarray}
where $i_{Kn},c_{Kn} \to i_n, c_n$ and we invoke Einstein summation for readability, and
where $V_0$, $h$, $v$, and $\hat{a}$ ($\hat{a}^\dagger$) respectively refer to the external potential energy, the one- and two-electron Hamiltonian
integrals, and electron annihilation (creation) operators.
The number of electrons in the impurity subspace, $N_{I_K}$, is computed by the trace of the trial 1-RDM in the impurity block,
\begin{eqnarray}
    N_{I_K} &=& \sum_{i_{Kn}}^{M_{I_K}} \{\gamma_\mathrm{tr}\}^{i_{Kn}}_{i_{Kn}} = N_\mathrm{mol} - \sum_{c_{Kn}}^{M_{C_K}} \{\gamma_\mathrm{tr}\}^{c_{Kn}}_{c_{Kn}},
    \label{N_I}
\end{eqnarray}
which must be an integer, since the 1-RDM of a single determinant is idempotent (i.e., has eigenvalues of 0 or 1).
As Fig.\ \ref{embedding_1rdm_fig} shows, the 1-RDM in the embedding basis is block-diagonal by construction between impurity and core,
so the eigenvalues (and therefore the traces) of the two blocks are necessarily integers.

The second-quantized form of the impurity Hamiltonian presented by Eqs.\ (\ref{H_I})--(\ref{H1_I}), with a guaranteed integer electron number given by Eq.\ (\ref{N_I}), 
is practically significant, because this form is particularly amenable to computation. Algorithms for zero-temperature \emph{ab initio} wave function methods
with number-preserving, second-quantized Hamiltonian operators of this form are mature in quantum chemistry.

There are a few different ways to obtain the bath orbitals which are mathematically equivalent to Eq.\ (\ref{single_determinant_schmidt}),
if the 1-RDM involved is idempotent. However, in the above review, we have chosen to present Eq.\ (\ref{single_determinant_schmidt}) specifically,
and we have also chosen not to simplify Eqs.\ (\ref{H0_I}) and (\ref{H1_I}) using the single-determinantal form of $|\Phi_{C_K}\rangle$, in order to
highlight the difficulty in generalizing DMET to multi-determinantal $|\Psi_\mathrm{tr}\rangle$. It appears that
one could perform the SVD in Eq.\ (\ref{single_determinant_schmidt})
and build the Hamiltonian in Eqs.\ (\ref{H_I})--(\ref{H1_I}) using any trial wave function or indeed any 1- and 2-RDM, but then
\begin{enumerate}
    \item the Fock space of bath orbitals would not generally contain the true many-body bath states, and
    \item the right-hand side of Eq.\ (\ref{N_I}) would not generally evaluate to an integer.
\end{enumerate}
The first problem simply represents an approximation, but the second is a severe practical difficulty if interface with existing quantum-chemistry programs is desired.
It is the single-determinantal form of $|\Phi_\mathrm{tr}\rangle$ that prevents these problems.

One might suspect that DMET therefore cannot be useful for chemical systems characterized by strong electron correlation, since
for such systems, by definition, no single-determinantal wave function can be even qualitatively accurate. Clearly, the literature
shows\cite{Knizia2012,Knizia2013,Wouters2016,Sandhoefer2016,Pham2018} that this is not necessarily the case. 
The trial wave function only needs to generate bath states which span the same space as the bath states which would
be obtained by Schmidt decomposition of the true, correlated wave function. It does not need to be \emph{accurate} in that space for
DMET to be accurate. However, as discussed in Sec.\ \ref{intro}, the literature record 
does not prove that DMET is robustly useful for strong correlation in general. 

\subsection{Schmidt decomposition of MC-SCF wave functions \label{schmidt_analysis}}

Some multi-determinantal wave functions also have structure that can be used to simplify Schmidt decomposition.
In particular, a CAS($N_A$,$M_A$) wave function, which is the standard way of modeling strong correlation in quantum chemistry,
describes a molecule as two unentangled subsystems:\ an active-space part, $A$, defined by a general correlated wave function, $|\Psi_A\rangle$,
involving $N_A$ electrons occupying $M_A$ ``active orbitals'' indexed as $a_n, n=1,2,\ldots$, and 
an ``external'' part, $X$, consisting of a single determinant, $|\Phi_X\rangle$, in the space of $N_X=N_\mathrm{mol}-N_A$ electrons occupying $M_X=M_\mathrm{mol}-M_A$
external orbitals indexed as $x_n$,
\begin{eqnarray}
    |\mathrm{CAS}\rangle &=& |\Psi_A\rangle \otimes |\Phi_X\rangle.
\end{eqnarray}

The two subsystems are unentangled because, in terms of excitations from a reference determinant, the wave function
contains neither any excitations between $|\Phi_X\rangle$ and $|\Psi_A\rangle$ nor any 
excitations spanning both. Thus, the coefficient tensor in term of active and external orbitals is factorizable,
\begin{eqnarray}
    C^{\vec{a}}_{\vec{x}} &=& A^{\vec{a}} X_{\vec{x}},
    \label{cas_tensor_factorize1}
\end{eqnarray}
which holds for any choice of active orbitals and any choice of external orbitals, so long as the two sets
remain non-overlapping. If we specify that the active and external orbitals are each separately localized in real space,
and we use these semi-localized orbitals to define fragments consisting of $M_{A_K}$ active and $M_{X_K}$ external orbitals,
\begin{eqnarray}
    \{|f_{Kn}\rangle\} &=& \{|a_{Kn}\rangle\} \cup \{|x_{Kn}\rangle\},
    \label{schmidt_cas_definition}
\end{eqnarray}
so that
\begin{eqnarray}
    \sum_{a_{Kn}}^{M_{A_K}} |\langle a_{Kn}|f_{Ln}\rangle|^2 = \sum_{f_{Ln}}^{M_{F_L}} |\langle a_{Kn}|f_{Ln}\rangle|^2 &=& \delta_{KL},
    \label{schmidt_cas_condition}
\end{eqnarray}
then we can write
\begin{eqnarray}
    C^{\vec{a}_K\vec{x}_K}_{\vec{a}_L\vec{a}_M\cdots\vec{x}_L\vec{x}_M\cdots} &=& 
    A^{\vec{a}_K}_{\vec{a}_L\vec{a}_M\cdots} X^{\vec{x}_K}_{\vec{x}_L\vec{x}_M\cdots},
\end{eqnarray}
or, more simply,
\begin{eqnarray}
    C^{\vec{f}_K}_{\vec{e}_K} &=& A^{\vec{f}_K}_{\vec{e}_K} X^{\vec{f}_K}_{\vec{e}_K}.
    \label{cas_tensor_factorize2}
\end{eqnarray}
This implies that, as long as fragments are chosen to satisfy Eq.\ (\ref{schmidt_cas_condition}),
the Schmidt decomposition of a CAS wave function can be carried out as two separate Schmidt decompositions
for $|\Psi_A\rangle$ and $|\Phi_X\rangle$. The overall bath states are direct products
of the two subsystem sets of bath states,
\begin{eqnarray}
    |B_K\rangle &\to& |B_K^{(A)}\rangle \otimes |\vec{b}_K^{(X)}\rangle \otimes |\Phi_{C_K}^{(X)}\rangle,
    \label{cas_mbb}
\end{eqnarray}
where the parenthetical superscripts $(X)$ and $(A)$ remind the reader whether the corresponding factor describes
electrons in external orbitals or in active orbitals.
The bath states from $|\Phi_X\rangle$ are simplified as per Eq.\ (\ref{singdet_mbb}), because $|\Phi_X\rangle$ is 
a single determinant, but the bath states from $|\Psi_A\rangle$ remain general.

The simplification of the bath states to the form of Eq.\ (\ref{cas_mbb}) is not yet
practical, if the goal is to facilitate CASSCF by breaking it into coupled impurity problems. One still has to
perform SVD of $A^{\vec{f}_K}_{\vec{e}_K}$, which grows exponentially with the size of the active space,
if not the size of the whole system. This makes Schmidt decomposition comparable in computational expense
to the CASSCF calculation itself, just as Schmidt decomposition of a general wave function without any approximation
is comparable in expense to a FCI calculation on the whole system. 

However, CASSCF can be approximated to simplify the determination of $|B_K^{(A)}\rangle$, without
losing its applicability to strongly-correlated systems. For our purposes, the obvious application is to divide the active space
into subspaces based on the real-space localization to fragments: $A_K$, $A_L$, etc. The simplest example
of such a wave function has the active subspaces unentangled to each other, which is what we call LAS,
\begin{eqnarray}
    |\mathrm{LAS}\rangle &=& \left(\bigotimes_K |\Psi_{A_K}\rangle \right) \otimes |\Phi_X\rangle,
    \label{las_definition}
\end{eqnarray} 
which has no intersubspace excitations and no connected excitations (in the sense of the linked cluster
theorem of many-body perturbation theory\cite{MarchYoungSampanthar_mbpt,ShavittBartlett_mbpt})
spanning two or more subspaces. The active-space part of a LAS wave function thus has a coefficient tensor that factorizes,
\begin{eqnarray}
    A^{\vec{a}} &=& A^{\vec{a}_K\vec{a}_L\cdots}=A^{\vec{a}_K}A^{\vec{a}_L}\cdots,
    \label{las_factorize1}
\end{eqnarray}
or, recalling Eq.\ (\ref{schmidt_cas_condition}),
\begin{eqnarray}
    A^{\vec{f}_K}_{\vec{e}_K} &=& A^{\vec{f}_K} A_{\vec{e}_K}.
    \label{las_factorize2}
\end{eqnarray}
Note that in order for the whole wave function to observe electron number symmetry (i.e., $\hat{N}|\mathrm{LAS}\rangle=N_\mathrm{mol}|\mathrm{LAS}\rangle$),
each active subspace must individually observe electron number symmetry, $\hat{N}|\Psi_{A_K}\rangle=N_{A_K}|\Psi_{A_K}\rangle$ with integer $N_{A_K}$.

The SVD of a coefficient tensor that obeys Eq.\ (\ref{las_factorize2}) is trivial: there is one nonzero singular value, equal to unity,
and the left and right singular vectors are just the left and right factors themselves. Thus, the Schmidt decomposition
generates only one bath state,
\begin{eqnarray}
    |B_K^{(A)}\rangle &=& \bigotimes_{L\neq K} |\Psi_{A_L}\rangle,
\end{eqnarray}
for the $K$th fragment. Combining this with Eq.\ (\ref{cas_mbb}) we see that the bath states for the whole LAS wave function are
\begin{eqnarray}
    |B_K\rangle &\to& \left(\bigotimes_{L\neq K} |\Psi_{A_L}\rangle\right) \otimes |\vec{b}_K^{(X)}\rangle \otimes |\Phi_{C_K}^{(X)}\rangle,
\end{eqnarray}
which, for a single fragment, differ from each other only in the bath-orbital part, $|\vec{b}_K^{(X)}\rangle$, just as do the bath states for a single determinant.
The implication is that Eqs.\ (\ref{single_determinant_schmidt})--(\ref{N_I}), for the Schmidt decomposition [Eq.\ (\ref{schmidt_general})]
and impurity projection [Eq.\ (\ref{H_I_K})] of a single determinant, also apply \emph{without modification} to the Schmidt
decomposition and impurity projection of a LAS wave function, provided that they are evaluated exactly as written and
that the choice of entangled local fragments observes Eq.\ (\ref{schmidt_cas_condition}). We say that overall,
the LAS wave function with appropriately-chosen fragments \emph{has the same Schmidt decomposition} as a single determinant.

\begin{figure}[tb]
\includegraphics[scale=1.0]{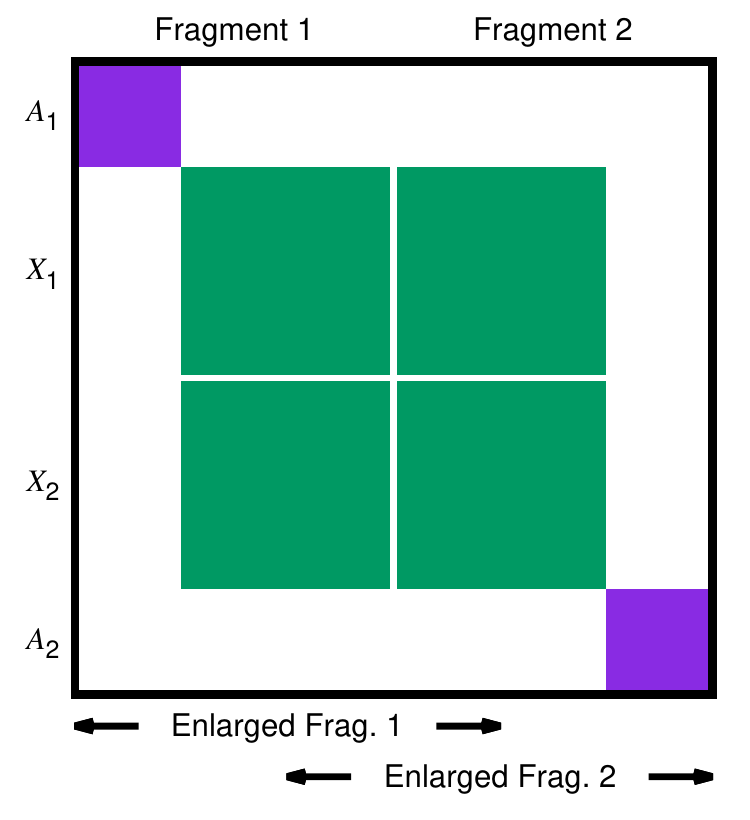}
\caption{Diagrammatic representation of the 1-RDM of a LAS wave function
with two active subspaces in the context of a LASSCF calculation with two fragments. Again, white space
denotes matrix elements that are identically zero. ``Enlarged fragments'' are discussed in Sec.\ \ref{enlarged_fragments} of
the Appendix.
\label{las_1rdm_fragbasis_fig}}
\end{figure}

\begin{figure}[tb]
\includegraphics[scale=1.0]{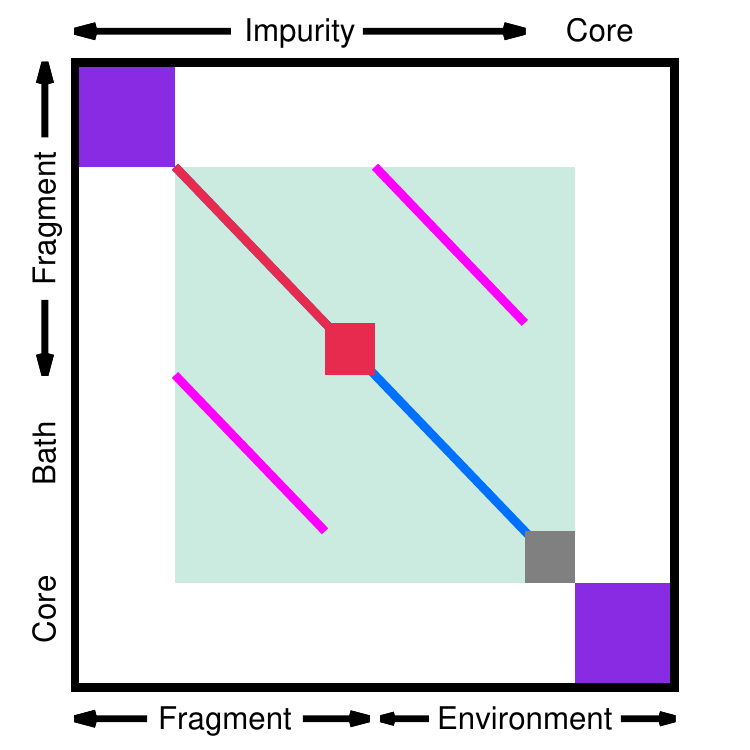}
\caption{The same as in Fig.\ \ref{las_1rdm_fragbasis_fig}, except in the embedding basis for one of the two identical fragments generated by the SVD
of the fragment-environment coupling block [Eq.\ (\ref{single_determinant_schmidt})]. White and light green space indicate matrix elements that are identically zero.
Light green space indicates external sector of orbitals, $X_1$ and $X_2$, represented by green blocks in Fig.\ \ref{las_1rdm_fragbasis_fig}. Compare
Fig.\ \ref{embedding_1rdm_fig}.
\label{las_1rdm_embbasis_fig}}
\end{figure}

Figures \ref{las_1rdm_fragbasis_fig} and \ref{las_1rdm_embbasis_fig} show this in terms of the 1-RDM of the LAS wave function.
Figure \ref{las_1rdm_fragbasis_fig} shows the 1-RDM in the fragment-orbital basis, and the fragment orbitals are selected in accordance
with Eq.\ (\ref{schmidt_cas_condition}). As with any active-space wave function, there are no nonzero off-diagonal elements coupling 
active and external orbitals. LAS also specifies that there is no entanglement between active subspaces, so there are no
nonzero elements coupling the two active subspaces either. The two fragments are entangled
solely through their external orbitals.

Figure \ref{las_1rdm_embbasis_fig} shows the effect of evaluating Eq.\ (\ref{single_determinant_schmidt}) using the 1-RDM of a LAS trial wave
function. The external orbitals are rotated among themselves, similar to what is depicted in Fig.\ \ref{embedding_1rdm_fig}, but
the active orbitals are not, because there is no entanglement of the active subsystems. Instead, the active-space part of fragment
$K$ is assigned in its entirety to the impurity, since it is contained in the fragment [Eq.\ (\ref{schmidt_cas_condition})],
and all other active subspaces are assigned, in their entireties, to the core. Since each active subspace describes an integer number
of electrons, the total number of electrons in the impurity [Eq.\ (\ref{N_I})] remains an integer. Therefore,
the numerical recipe of Eqs.\ (\ref{single_determinant_schmidt})--(\ref{N_I}) applied to a LAS wave function
generates an impurity subsystem with the same attractive compatibility
with standard quantum-chemistry approximations as a single determinant. 

This suggests an obvious method for variational optimization of a LAS wave function by adapting 
the algorithm of standard DMET. This is the method we name LASSCF. In essence, it exploits the fact that the Schmidt
decomposition of a LAS wave function is much less costly than variationally obtaining a LAS wave function otherwise would be, and
that therefore cycling through Eqs.\ (\ref{schmidt_general})--(\ref{impurity_problem_nomu}) actually provides a practical advantage,
unlike the case with a general (FCI) or CAS $|\Psi_\mathrm{tr}\rangle$.
The Schmidt decomposition of a LAS $|\Psi_\mathrm{tr}\rangle$ defines a CASSCF impurity problem for each fragment,
which can be solved with a standard implementation of CASSCF. The active orbitals and CI vectors
are collected to generate an update of the trial wave function, and the process is repeated until the trial wave function converges.
No correlation potential or chemical potential is required because the embedding is exact.

The catch is that in each iteration,
the active orbitals are shifted by the CASSCF impurity calculation in a way that generally does not respect Eq.\ (\ref{schmidt_cas_condition}).
This is essential, because we cannot precisely know the shapes of the optimized active orbitals ahead of time.
Fortunately, nothing obligates a DMET-like method to use a single set of fragment orbitals chosen by the user
at the beginning of the calculation. Instead, in LASSCF, the user provides an initial guess for a set of localized fragment orbitals.
In successive iterations, these fragment orbitals ``relax:'' they shift to repeatedly restore Eq.\ (\ref{schmidt_cas_condition}),
while remaining as close as possible to the initial guess.
We discuss how to accomplish this, and how to initialize the self-consistent
cycle, in the next section.

\section{Localized active-space self-consistent field technical aspects\label{lasscf}}

To summarize Sec.\ \ref{schmidt_analysis}:
LASSCF refers to the variational optimization of a LAS trial wave function, 
defined by Eq.\ (\ref{las_definition}), where the product subscript $K$ is over localized,
entangled fragment subsystems which tile the molecule and in which the unentangled active subsystems, $A_K$, are confined.
The optimization is carried out by modifying the algorithm of DMET; i.e., by cycling through Eqs.\ (\ref{schmidt_general})--(\ref{impurity_problem_nomu})
with $|\Psi_\mathrm{tr}\rangle=|\mathrm{LAS}\rangle$ until self-consistency, while exploiting the form of the LAS wave function to simplify the first step, the Schmidt decomposition.
This leads to the optimization of the active orbital coefficients within separate impurity problems spanning various semi-localized orbital subspaces,
which existing implementations of CASSCF can solve without modification.
The impurity problems are coupled to each other through the values of the effective Hamiltonian integrals [Eqs.\ (\ref{H_I})--(\ref{H1_I})]
and the mutual orthogonality constraint [Eq.\ (\ref{schmidt_cas_condition})].

LASSCF resembles DMET in that LAS has the same Schmidt decomposition as a single determinant given an appropriate choice of fragments,
and therefore much of the DMET algorithm can be imported into the LASSCF algorithm. They differ in that LASSCF is always variational
and provides a wave function, and instead of a chemical potential or a correlation
potential, the definition of the fragments, $|f_{Kn}\rangle$, is shifted over subsequent iterations.
The DMET-like algorithm also means that LASSCF is practical for large systems, since it efficiently breaks
a formally large MC-SCF problem into several small coupled MC-SCF problems. The LASSCF energy is an upper bound
for the corresponding CASSCF energy.

After defining the molecule and the atomic orbital (AO) basis set, both DMET and LASSCF must first transform
the AO basis into a form that is both orthonormal and localized, using, for example, the Foster-Boys\cite{Foster1960}
or ``meta-L\"{o}wdin''\cite{Sun2014,Sun2018} method.
%We use the method of Sun and Chan\cite{Sun2014} 
%which orthonormalizes basis functions without sacrificing their atomic-orbital character, which
%is implemented in the {\sc PySCF}\cite{Sun2018} package as ``meta-L\"{o}wdin orthogonalization.''
Then, the LASSCF self-consistent cycle is
\begin{enumerate}
    \item Set $|g_{Kn}\rangle, N_{A_K}, M_{A_K}$; initial guess of $|\Psi_\mathrm{tr}\rangle$; $|g_{Kn}\rangle \to |f_{Kn}\rangle$,
    \item For all fragments (i.e., for all $K$):
    \begin{enumerate}
        \item Schmidt decomposition, impurity projection,
        \item CASSCF calculation: get updated $|a_{Kn}\rangle$, $\gamma_{A_K}$, $E_{I_K}$,
    \end{enumerate}
    \item Update $|\Psi_\mathrm{tr}\rangle$, $|f_{Kn}\rangle$,
    \item If not converged, return to step 2,
\end{enumerate}
where $|g_{Kn}\rangle$ refers to ``guess fragment orbitals'' serving as an initial approximation
to $|f_{Kn}\rangle$, the ``fragment orbitals'' that shift repeatedly to observe Eq.\ (\ref{schmidt_cas_condition}), and where
$\gamma_{A_K}$ refers to the 1-RDM and 2-RDM of the active-space part of the $K$th fragment.

The Schmidt decomposition was explored in detail in Secs.\ \ref{dmet_background} and \ref{schmidt_analysis}, and the CASSCF
impurity calculation is standard. Therefore, in this section, we describe only essential technical details related to initialization and wave function and fragment
orbital updating. Non-essential technical details relating to updating the fragment orbitals
and details concerning the initialization of the CASSCF impurity calculation (which apply to both CAS-DMET and LASSCF)
are discussed in Secs.\ \ref{enlarged_fragments} and \ref{impsolve_guesses} of the Appendix, respectively.

Although the theoretical development of LASSCF was presented in terms of spinorbitals for the sake of conceptual simplicity,
our implementation assumes spin-symmetric spatial orbitals, and Eqs.\ (\ref{single_determinant_schmidt})--(\ref{N_I}) are evaluated in terms of spin-summed
density matrices.

\subsection{LASSCF initialization\label{guesses}}

The user supplies the active subspace for each fragment, $N_{A_K}$ electrons in $M_{A_K}$ orbitals;\ an initial guess
for the whole-molecule trial wave function, $|\Psi_\mathrm{tr}\rangle$;\ and the guess fragment orbitals, $|g_{Kn}\rangle$.
The guess fragment orbitals can be constructed in the same way that fragment orbitals are constructed in DMET.
For example, if meta-L\"{o}wdin orthogonalization\cite{Sun2014,Sun2018} is used and the molecule is an iron complex, then the orthogonalized
AOs of the central iron atom can be defined as $\{|g_{11}\rangle, |g_{12}\rangle,\ldots\}$, the orthogonalized AOs of one ligand can be defined as $\{|g_{21}\rangle, |g_{22}\rangle, \ldots\}$, 
another ligand as $\{|g_{31}\rangle, |g_{32}\rangle, \ldots\}$, etc.
These guesses are not discarded after the first iteration - we make use of them throughout the calculation in order to keep
the fragment orbitals ($|f_{Kn}\rangle$) local in real space (see Secs.\ \ref{update_frag} and \ref{enlarged_fragments} below). 
The RHF wave function serves as an adequate initial guess of $|\Psi_\mathrm{tr}\rangle$ in the calculations reported in this work.

\subsection{Updating the trial wave function\label{update_las}}

The impurity calculations relax the active orbitals from each active subspace independently and so, in general, cause them to overlap,
\begin{eqnarray}
    \sum_{L\neq K}\sum_{a_{Ln}}^{M_{A_L}} |\langle a_{Kn}|a_{Ln}\rangle|^2 > 0.
\end{eqnarray}
This must be corrected at this stage.
We carry out L\"{o}wdin orthogonalization of the overlapping $|a_{Kn}\rangle$ to generate orthogonal
whole-molecule active orbitals, $|a_n\rangle$.
In the new orthogonal basis, we sum the RDMs from the active subspaces to generate RDMs for the whole-molecule active space,
\begin{eqnarray}
    \{\gamma_\mathrm{tr}\}^{a_1}_{a_2} &=& \sum_K \{\gamma_{A_K}\}^{a_1}_{a_2},
    \label{gamma_A}
    \\
    \{\lambda_\mathrm{tr}\}^{a_1a_3}_{a_2a_4} &=& \sum_K \{\lambda_{A_K}\}^{a_1a_3}_{a_2a_4},
    \label{lambda_A}
\end{eqnarray}
where $\lambda$ refers to the cumulant expansion of the RDMs,\cite{Kutzelnigg1999}
\begin{eqnarray}
    \gamma^{a_1a_3}_{a_2a_4} &=& \lambda^{a_1a_3}_{a_2a_4} + \gamma^{a_1}_{a_2}\gamma^{a_3}_{a_4}
    - \gamma^{a_1}_{a_4}\gamma^{a_3}_{a_2}.
    \label{lambda}
\end{eqnarray}

Whole-molecule external orbitals, $|x_n\rangle$, are constructed by diagonalizing the projection operator
\begin{eqnarray}
    \hat{P}_{X_\mathrm{mol}} &\equiv& \sum_{x_n} |x_n\rangle\langle x_n|,
    = 1-\sum_{a_n}|a_n\rangle\langle a_n|.
    \label{P_X_mol}
\end{eqnarray}
The molecular Hamiltonian is then projected into the external space. 
This corresponds to Eqs.\ (\ref{H_I})--(\ref{H1_I}), with the
substitutions $I_K\to X_\mathrm{mol}$, $i_n\to x_n$, and $c_n\to a_n$ throughout. A HF calculation using this projected Hamiltonian 
for $N_{X_\mathrm{mol}}=N_\mathrm{mol}-\sum_K N_{A_K}$ electrons produces the whole-molecule external determinant,
$|\Phi_X\rangle$, which gives the external part of the updated trial 1-RDM used
for the Schmidt decompositions [Eq.\ (\ref{single_determinant_schmidt})] in the next iteration.

\subsection{Updating the fragment orbitals \label{update_frag}}

After updating the trial wave function, the next iteration's fragment orbitals are constructed,
\begin{eqnarray}
    \{|a_{Kn}\rangle\} \cup \{|x_{Kn}\rangle\} \to \{|f_{Kn}\rangle\},
\end{eqnarray}
where $|a_{Kn}\rangle$ ($|x_{Kn}\rangle$) is a linear combination of various whole-molecule active (external) orbitals, $|a_n\rangle$ ($|x_n\rangle$) from Sec.\ \ref{update_las}.
We find that the active orbital overlap induced by orbital optimization in separate impurities is small, so that orthogonalization changes little and it is
straightforward to assign each orthonormal active orbital to one fragment, $|a_n\rangle\to |a_{Kn}\rangle$, without further transformation. As for
the external orbitals, our implementation approximately corresponds to solving
\begin{eqnarray}
    \hat{P}_{X_\mathrm{mol}}\hat{P}_{G_K}|x_{Kn}\rangle &=& \lambda_{x_{Kn}}|x_{Kn}\rangle,
    \label{assign_external}
\end{eqnarray}
where
\begin{eqnarray}
    \hat{P}_{G_K} &\equiv& \sum_{g_{Kn}}^{M_{F_K}} |g_{Kn}\rangle\langle g_{Kn}|,
    \label{P_G_K}
\end{eqnarray}
and retaining the $M_{X_K}$ highest-eigenvalue solutions of Eq.\ (\ref{assign_external}) as external orbitals of the $K$th fragment.
This helps to keep the active orbitals local in real space by providing a semi-local set of orbitals in which to relax
during the next orbital-optimization step, inasmuch as $|g_{Kn}\rangle$ are chosen
by the user to be local. The specifics, however, are relegated to Sec.\ \ref{enlarged_fragments} in the Appendix.

Orthogonalization of the active orbitals generates small nonzero 1-RDM elements coupling the active subspaces, $\{\gamma_\mathrm{tr}\}^{a_{Kn}}_{a_{Ln}}$,
which go to zero as LASSCF converges.
In practice, we ignore these by just projecting $\gamma_\mathrm{tr}$ into the external space (i.e., the light green
region of Fig.\ \ref{las_1rdm_embbasis_fig}) before evaluating Eq.\ (\ref{single_determinant_schmidt}). More importantly, it may cause the number of electrons
in each active subspace to deviate from an integer. The electron number error must be evaluated,
\begin{eqnarray}
    \epsilon_{A_K} &\equiv& N_{A_K} - \sum_{a_{Kn}}^{M_{A_K}} \{\gamma_\mathrm{tr}\}^{a_{Kn}}_{a_{Kn}}.
    \label{eps_A_K}
\end{eqnarray}
In our implementation, an error is raised if any $\epsilon_{A_K}$ is greater in magnitude than $10^{-2}$.
This never happens in our calculations reported below, and $\epsilon_{A_K}$ also always goes to zero as LASSCF converges.

\section{Test calculations \label{results_discussion}}

% figure* for column-spanning form in two-column format
\begin{figure*}[tb]
%\begin{figure}[tb]
\includegraphics[width=1.0\textwidth]{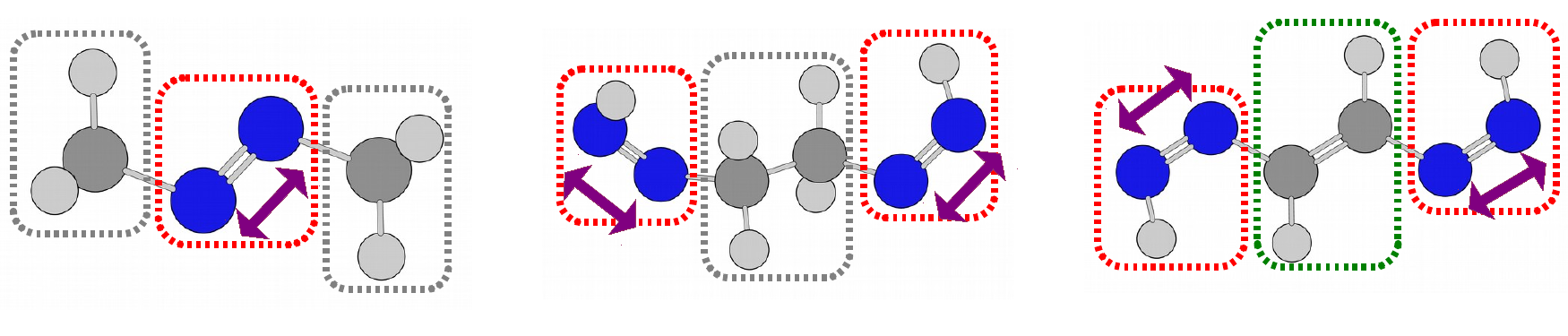}
\caption{Three molecules whose nitrogen-nitrogen bond dissociation (purple arrows) potential energy curves were examined
by LASSCF in this work: azomethane (left), C$_2$H$_6$N$_4$ (center), and C$_2$H$_4$N$_4$ (right), depicted at their B3LYP/6-31g(d,p)
optimized geometries.
The fragments in DMET and the guess fragments in LASSCF used in our calculations are defined by the orthogonalized atomic orbitals
of sets of atoms enclosed by dashed rectangles. Red dashed rectangles indicate impurities with an active space of (4,4), grey dashes
indicated (0,0), and green dashes indicate that both (0,0) and (2,2) were tested. For the two bisdiazenes, the potential energy surface
was scanned along simultaneous stretching of both nitrogen-nitrogen bonds.
\label{azene_fig}}
\end{figure*}
%\end{figure}

All equilibrium molecular geometries in this work were obtained at the B3LYP/6-31g(d,p)
level of theory using Gaussian 09,\cite{g09_e01} and the 6-31g basis set was used in subsequent potential energy scans.
{\sc PySCF}\cite{Sun2018} was used for the whole-molecule CASSCF calculations.
Our implementation\cite{mrh_software} of both DMET and LASSCF is a local modification of the fork\cite{pham_software}
of the QC-DMET package\cite{qcdmet_software} that was used to perform the calculations reported in Ref.\ \citenum{Pham2018}.

We use the terminology of Ref.\ \citenum{Pham2018} and that discussed in Sec.\ \ref{dmet_background}
in describing variants of DMET examined in this section. ``scCAS-DMET'' refers to a DMET calculation 
utilizing CASSCF as the impurity solver,
in which the correlation potential, $\{\hat{u}_{K}\}$ of Eq.\ (\ref{trial_eigenequation}), is fully optimized
by minimizing the error function
\begin{eqnarray}
    \epsilon(\{\hat{u}_K\}) &\equiv& \sum_K\sum_{i_{K1}\le i_{K2}} \left|\{\gamma_{I_K}\}^{i_{K1}}_{i_{K2}}-\{\gamma_\mathrm{tr}\}^{i_{K1}}_{i_{K2}}\right|^2.
    \label{errfunc}
\end{eqnarray}
``CAS-DMET'' refers to a ``one-shot'' approximation in which the correlation potential is dropped
and $|\Phi_\mathrm{tr}\rangle$ is taken to be the RHF wave function.
Most calculations reported here use the ``democratic'' form of DMET, but in Sec.\ \ref{results_azomethane} below we also explore
the ``single-embedding'' variant.

We tested LASSCF's performance in replicating comparable CASSCF
potential energy curves of the three molecules depicted in Fig.\ \ref{azene_fig}: azomethane, 2-diazenylethyldiazene (hereafter C$_2$H$_6$N$_4$),
and 2-diazenylethenyldiazene (hereafter C$_2$H$_4$N$_4$).
We scan the potential energy surface along the N$_2$ bond length coordinate. For the two bisdiazenes, we scan along the simultaneous stretching
of both bonds;\ at the equilibrium geometries, they have the same length due to point-group symmetry. We split each of the three molecules
into three fragments in our DMET and LASSCF calculations, as depicted in Fig.\ \ref{azene_fig}, and assign those fragments containing an N$_2$
an active space of (4,4) in the CASSCF impurity calculations. To the remaining fragments we assign an active space of (0,0)
for the azomethane and C$_2$H$_6$N$_4$ molecules, and either (0,0) or (2,2) for the C$_2$H$_4$N$_4$ molecule.
Most LASSCF calculations reported below converged in their total energies to a tolerance of $10^{-8}\ E_\mathrm{h}$ within the first five iterations.

\subsection{Azomethane \label{results_azomethane}}

\begin{figure}[tb]
\includegraphics[scale=1.0]{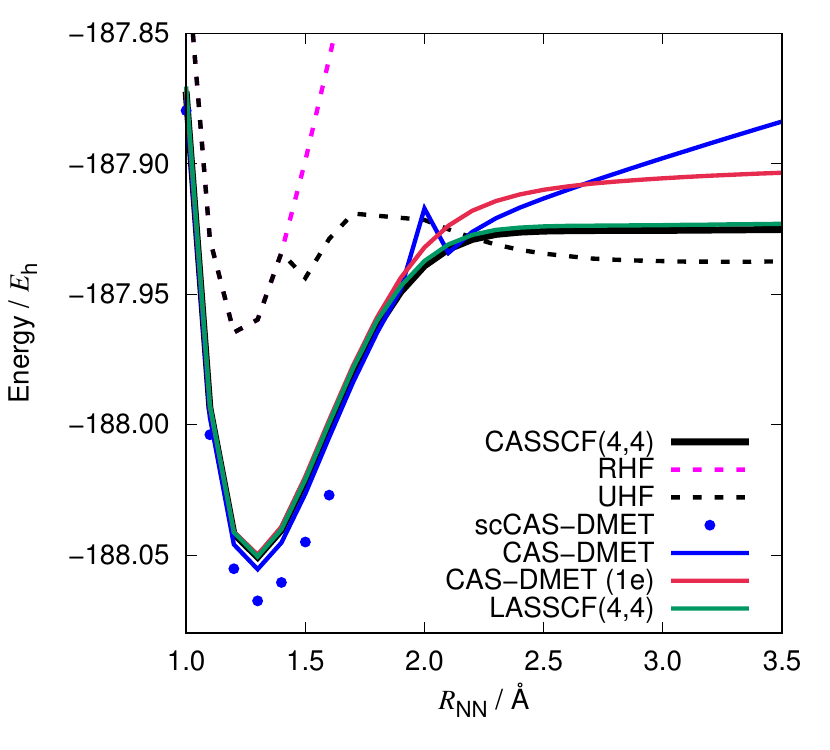}
\caption{Potential energy surface of the azomethane molecule along the nitrogen-nitrogen bond-stretching coordinate,
calculated with various methods. In the legend, ``CAS-DMET (1e)'' refers to a ``single-embedding'' calculation (see text).
The optimization of the correlation potential in the scCAS-DMET calculations failed to converge for $R_\mathrm{NN}>1.6\ \AA$.
\label{me2n2_pes_fig}}
\end{figure}

\begin{figure}[tb]
\includegraphics[scale=0.9]{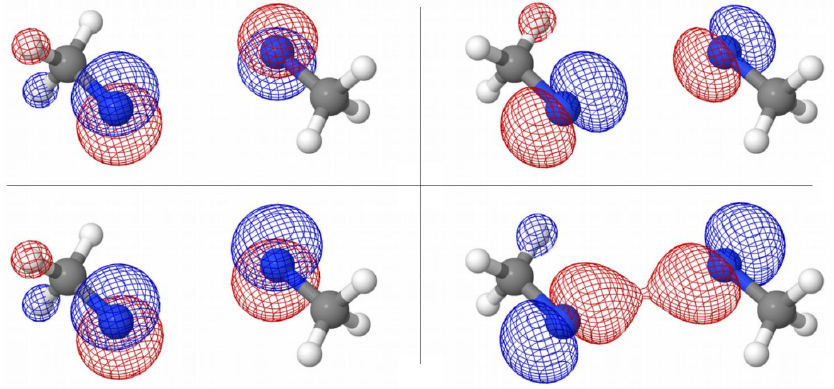}
\caption{The four most strongly-entangled bath orbitals of the impurity for the right methyl fragment in the CAS-DMET calculation of
azomethane at $R_\mathrm{NN}=3.5\ \AA$.
\label{bad_entangle_fig}}
\end{figure}

Figure \ref{me2n2_pes_fig} shows the potential energy curve of N$_2$ stretching calculated with RHF, spin-unrestricted Hartree--Fock (UHF), CASSCF(4,4),
CAS-DMET, scCAS-DMET, single-embedding CAS-DMET, and LASSCF(4,4). The CASSCF(4,4) curve was demonstrated in Ref.\ \citenum{Pham2018} to be qualitatively
converged with respect to the size of the active space, as chemical intuition suggests for the breaking of a double bond in a simple molecule such as azomethane.
The CAS-DMET results are as in Ref.\ \citenum{Pham2018}: CAS-DMET follows the CASSCF curve until $R_\mathrm{NN}= 2\ \AA$, and
then fails to plateau at the dissociation energy, resembling the behavior of RHF under covalent bond breaking in that the total energy continually increases.

Figure \ref{me2n2_pes_fig} also reports the results of single-embedding CAS-DMET and scCAS-DMET calculations of the azomethane potential energy curve.
Single-embedding CAS-DMET was shown in Ref.\ \citenum{Pham2018} to give results nearly identical to CASSCF(4,4) for the similar pentyldiazene molecule.
Here, unlike democratic CAS-DMET, the single-embedding calculations appear to succeed in breaking the nitrogen double bond, but they also
predict a dissociation energy at least 20 m$E_\mathrm{h}$ higher than CASSCF(4,4).
Meanwhile, the scCAS-DMET calculations actually fail outright for $R_\mathrm{NN}>1.6\ \AA$:\ the correlation potential fails to reach a stationary point.
The onset of this failure is at a shorter bond distance than the visible separation of the CASSCF and CAS-DMET curves
in Fig.\ \ref{me2n2_pes_fig}, but it is close to the point at which spontaneous symmetry breaking at the mean-field level occurs,
as shown by the splitting of the UHF and RHF curves. 

By contrast, the LASSCF potential energy curve is nearly indistinguishable from the CASSCF curve on this scale;\ numerically, the LASSCF
dissociation energy is 2 m$E_\mathrm{h}$ higher than the CASSCF prediction. What causes DMET to encounter such difficulties in this system, compared to LASSCF?
Reference \citenum{Pham2018} reported that truncating the bath; i.e., removing all but a small number of the most strongly-entangled bath orbitals from the impurity subspace
of the N$_2$ fragment, improves the qualitative shape of the CAS-DMET curve. This implies that the trial wave function is providing a bad bath space to the impurity. But why
then do we not, in the scCAS-DMET calculations, find a correlation potential that solves this problem?

The orbitals depicted in Fig.\ \ref{bad_entangle_fig} closely resemble the CASSCF(4,4) optimized active orbitals of the N$_2$ fragment:\ $\sigma$ and $\pi$ bonding and antibonding
linear combinations of nitrogen valence orbitals, spanning the two bonds that are broken. However, they are in fact the four most strongly-entangled [i.e., largest
singular values of Eq.\ (\ref{single_determinant_schmidt})] \emph{bath orbitals} for the methyl fragment on the right in the CAS-DMET calculation at $R_\mathrm{NN}=3.5\ \AA$. 
The methyl impurity problems are solved with RHF, and since the methyl impurity contains these broken-bond orbitals, the inappropriate single-determinantal model
of bond breaking affects the overall correlation energy \emph{via} Eqs.\ (\ref{1RDM_output}) and (\ref{2RDM_output}). 

These bath orbitals are pathological, assuming dynamical correlation doesn't qualitatively change the CASSCF(4,4) picture of the wave function.
The active space of a CAS wave function has \emph{zero} entanglement to any orbitals
in the external space, as discussed in Sec.\ \ref{schmidt_analysis}. 
The pathological entanglement of the nitrogen active orbitals to the methyl atomic orbitals corresponds to nonzero off-diagonal Fock
matrix elements coupling these degrees of freedom. In principle, this could be prevented with a correlation potential that cancels these off-diagonal Fock matrix terms.
However, the DMET correlation potential cannot do this, because these pathologically-entangled degrees of freedom are on separate fragments, and the DMET
correlation potential is local.

LASSCF does not have any such difficulty. LASSCF has exact embedding
with a qualitatively correct form of the wave function, and the only guesses that the user must provide (fragment choice, active subspace choice, and active orbital initialization) correspond
to user-supplied parameters that are also required by standard DMET. LASSCF also has no computational step with a higher cost scaling than standard DMET. In fact,
for these small systems, LASSCF is significantly faster than the successful scCAS-DMET calculations; for the latter, at least in our implementation, the slowest step was the correlation-potential minimization problem.

\subsection{Bisdiazenes}

\begin{figure}[tb]
\includegraphics[scale=1.0]{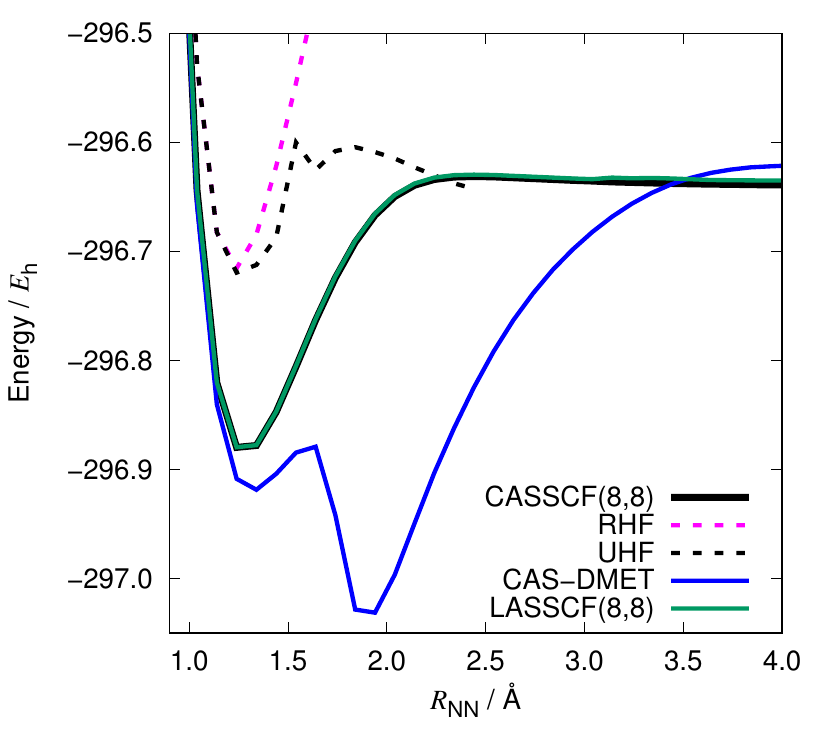}
\caption{Potential energy surface of the C$_2$H$_6$N$_4$ molecule along the simultaneous nitrogen-nitrogen bonds-stretching coordinate,
calculated with RHF, UHF, CASSCF, CAS-DMET, and LASSCF. UHF calculations proved difficult to converge consistently for $R_\mathrm{NN}>2.4\ \AA$
so this data is not shown.
\label{c2h6n4_pes_fig}}
\end{figure}

\begin{figure}[tb]
\includegraphics[scale=1.0]{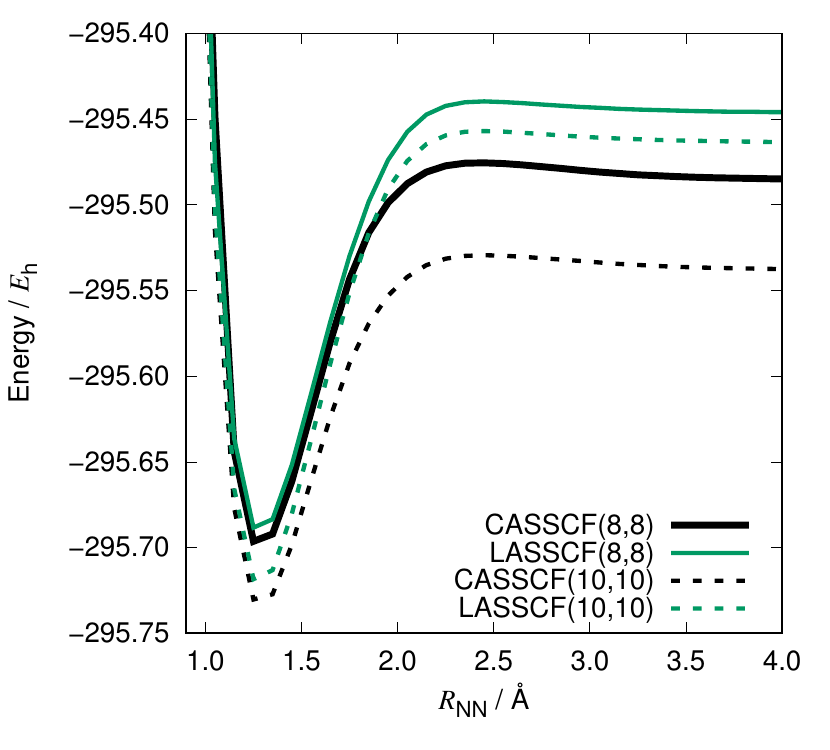}
\caption{Potential energy surface of the C$_2$H$_4$N$_4$ molecule along the simultaneous nitrogen-nitrogen bonds-stretching coordinate,
calculated with CASSCF and LASSCF with the (8,8) and (10,10) active spaces.
\label{c2h4n4_pes_fig}}
\end{figure}

\begin{figure*}[tb]
%\begin{figure}[tb]
\includegraphics[width=1.0\textwidth]{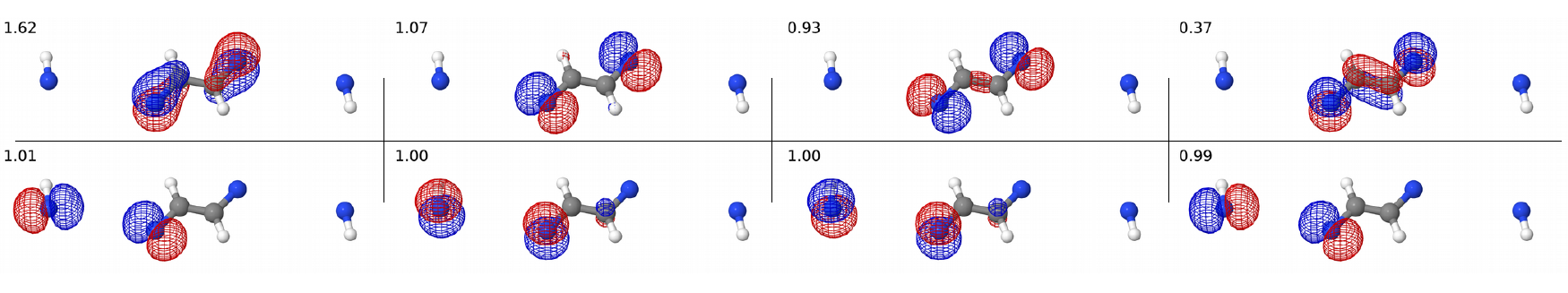}
\caption{Four natural orbitals of the C$_2$H$_4$N$_4$ molecule at $R_\mathrm{NN}=4.25\ \AA$ and their occupancies, calculated
with CASSCF(8,8) (top) and LASSCF(8,8) (bottom). The remaining four CASSCF(8,8) natural active orbitals are the $p_x$ and $p_y$
orbitals of the nitrogen atoms of the dissociated NH diradicals, each with an occupancy of 1.00. The remaining four LASSCF(8,8) natural active orbitals
are equivalent to those displayed here under a C$_2$ rotation of the molecule.
\label{c2h4n4_nos_fig}}
\end{figure*}
%\end{figure}

Single-embedding DMET performs somewhat better than CAS-DMET for azomethane and is nearly identical to CASSCF in the pentyldiazene calculations reported
in Ref.\ \citenum{Pham2018}. However, it is inapplicable to systems with multiple strongly-correlated fragments, such as our C$_2$H$_6$N$_4$ and C$_2$H$_4$N$_4$
bisdiazene molecules with the two nitrogen bonds simultaneously broken. We expect LASSCF to perform well for such molecules
if the active subspaces on adjacent fragments aren't strongly entangled, as anticipated for C$_2$H$_6$N$_4$. On the other hand,
the C$_2$H$_4$N$_4$ molecule has a $\pi$-bond system connecting the two diazene groups,
and chemical intuition suggests that as the bonds break, $\pi$ electrons on the central two nitrogens will recouple through the central C$_2$H$_2$ unit.
We therefore compare the performance of LASSCF for the potential energy curves of these two systems to investigate how the accuracy of the LAS wave function is affected
by this simple example of possible intersubspace entanglement.

Figure \ref{c2h6n4_pes_fig} shows the potential energy curves for C$_2$H$_6$N$_4$, calculated with CASSCF(8,8), LASSCF(8,8), and CAS-DMET, as well as RHF and UHF
in order to once again highlight the putative onset of strong electron correlation. CAS-DMET's performance, in terms of replicating the CASSCF curve, is very
poor and this curve is (probably) completely unphysical in the region where mean-field spin-symmetry breaking is significant ($R_\mathrm{NN}>\ 1.6\ \AA$). Democratic
DMET is not variational, unlike single-embedding DMET or LASSCF, so the lower energy predicted by CAS-DMET for $1.2\ \AA\le R_\mathrm{HH}\le 3.5\ \AA$
does not suggest superiority to CASSCF. We also note that it was significantly more difficult to 
generate the CAS-DMET curve in Fig.\ \ref{c2h6n4_pes_fig} than it was to generate the LASSCF curve, in that the former was much more sensitive
to the quality of the initial guess active orbitals than the latter.

For C$_2$H$_6$N$_4$, LASSCF again follows the CASSCF curve closely enough to be indistinguishable on the scale of Fig.\ \ref{c2h6n4_pes_fig};\ the difference
in dissociation energies is 4 m$E_\mathrm{h}$.
In this case, we expect the active subspaces not to be strongly entangled, so this result is unsurprising. On the other hand,
for C$_2$H$_4$N$_4$, the disagreement between LASSCF and CASSCF at dissociation is closer to 38 m$E_\mathrm{h}$ for the (8,8) active space
and 74 m$E_\mathrm{h}$ for the (10,10) active space. The constraint of unentangled subspaces in LAS wave functions seems to be more severe
for this system due to the coupling of the active electrons through the central double bond, although it may be that a different
partition of the (10,10) overall active space to the three fragments would improve the LASSCF results.

Figure \ref{c2h4n4_nos_fig} displays several natural orbitals and their occupancies of C$_2$H$_4$N$_4$ at dissociation calculated with LASSCF(8,8) and CASSCF(8,8).
The LASSCF calculation breaks the double bonds completely and generates an octo-radical species, where the $p_x$ and $p_y$ atomic orbitals of each nitrogen atom are singly occupied.
This is similar to the picture that both LASSCF and CASSCF draw for the C$_2$H$_6$N$_4$ molecule at dissociation (not pictured), as well as the weakly-recoupled electrons
of the broken $\sigma$ bonds for the C$_2$H$_4$N$_4$ CAS wave function (center top of Fig.\ \ref{c2h4n4_nos_fig}), whose natural-orbital occupancies
differ from unity by less than 0.1. But the electrons of the broken $\pi$ bonds of CASSCF C$_2$H$_4$N$_4$ (top left and right of Fig.\ \ref{c2h4n4_nos_fig}) recouple strongly,
leading to natural orbital occupancies of 1.62 and 0.37 for the $\pi$-orbital with one node and two nodes, respectively. LASSCF cannot replicate this recoupling because of the
product form of the wave function [Eq.\ (\ref{las_definition})].
The (10,10) active space does not qualitatively change this picture;\ the extra two active orbitals on the central fragment in LASSCF(10,10) correspond to the zero-node and three-node
$\pi$ orbitals (not pictured), with natural occupancies of close to 2 and 0, respectively.

Despite the clear problems with this LASSCF picture of the dissociation of the C$_2$H$_4$N$_4$ molecule, the LASSCF potential energy curves of Fig.\ \ref{c2h4n4_pes_fig} have 
good overall shape, and the quantitative disagreement in dissociation energies is of a reasonable magnitude. While, again, it may be that systematic exploration
of different options for partitioning the active space to the three fragments leads to superior agreement with CASSCF, it may also prove that perturbative correction to the current LASSCF
depiction of C$_2$H$_4$N$_4$ by the introduction of CASPT2\cite{Andersson1992} or MC-PDFT\cite{Manni2014} corrections
is sufficient to bring LAS-based results into close agreement with comparable
full-molecule calculations.

\section{Conclusions and future work\label{conclusions}}

DMET is a fully quantum-mechanical embedding method that has shown great promise in applicability to strongly-correlated molecules and materials
in tests on several simple model systems. However, we have seen that the good results for model systems do not necessarily generalize to more
realistic depictions of actual molecules. In revisiting the case of the potential energy curve of nitrogen bond breaking of azomethane, first
explored in Ref.\ \citenum{Pham2018}, we have shown evidence that the local correlation potential is not capable of 
modifying a single-determinantal trial wave function to produce a qualitatively good picture of entanglement, with significant consequences
for the accuracy of DMET calculations.

Of course, we have carried out this analysis in the context of the use of CASSCF as the impurity solver, and these conclusions
do not necessarily generalize to the use of FCI for the impurity subsystem problems, which is the more common context in which DMET is invoked.
This is because our interest is in uniting real-space embedding techniques with MC-SCF calculations, and in the context in which
MC-SCF calculations are typically necessary, the single determinant that DMET relies upon for its bath orbitals cannot be a qualitatively
good picture of the system. However, we have shown that the DMET algorithm can be generalized to use a particular type of active-space wave function,
LAS, to generate bath states with very little modification. The resulting method, LASSCF, features exact embedding and so lacks
the need for a chemical potential, correlation potential, or ambiguous choice of error function which are all artifacts of the mismatch
between the trial wave function and the high-level method one is actually attempting to use in standard DMET.

The cost of interface with the DMET algorithm for LASSCF is that it significantly constrains the CI vectors for the active space, more so
than even a GAS wave function with ``disconnected'' subspaces (i.e., each subspace has fixed electron number).
The latter may still have \emph{linked} terms between subspaces in the coupled-cluster expansion of its CI vector, but in LAS,
the active subspace parts of the wave function are constrained to a product form. Our initial numerical tests show that LASSCF results
are indistinguishable from CASSCF in situations in which it is various LAS wave functions are reasonably unentangled, as in, for example, the C$_2$H$_6$N$_4$ system where the two
N=N double bonds are separated by a single C-C bond. In the C$_2$H$_4$N$_4$ case, on the other hand, where there are three consecutive double bonds, the wave function
cannot be easily separated into localized, unentangled parts. However, in the latter case the LAS wave function still gives a reasonable N-N dissociation curve.

Moreover, the LAS wave function is an interesting concept in itself that can be explored also in conventional quantum mechanical calculations that do not require an embedding scheme.
The computational cost of conventional CASSCF, RASSCF, or GASSCF calculations grows exponentially with respect to the size of the active space
of the entire system. LAS, on the other hand, due to its localized product-form wave function, has an exponential cost scaling only with the growth of the individual fragment active subspaces.

% acschemo: command
% revtex: environment

\acknowledgement
%\begin{acknowledgments}
We thank Sina Chiniforoush, Christopher J. Cramer, Riddhish Umesh Pandharkar, Hung Pham, and Donald G. Truhlar
for useful discussion. This work was supported by the U.S. Department of Energy, Office of Basic Energy Sciences,
Division of Chemical Sciences, Geosciences and Biosciences under Award No.\ DEFG02-12ER16362.
%\end{acknowledgments}

\begin{suppinfo}
Absolute electronic energies, equilibrium molecular geometry Cartesian coordinates, and selected molecular orbital data.
\end{suppinfo}

\appendix

\section{Additional technical aspects of LASSCF implementation}

\subsection{Selection of external fragment orbitals\label{enlarged_fragments}}

In LASSCF, the selection of fragment orbitals must obey Eq.\ (\ref{schmidt_cas_condition}), which
implies that the active orbitals, $|a_{Kn}\rangle$, are fully orthonormal across all fragments.
However, there is no such restriction for the external fragment orbitals, $|x_{Kn}\rangle$.
In standard DMET, in the democratic\cite{Wouters2016} formalism, the only reason fragments
must be non-overlapping is ensure the applicability of Eqs.\ (\ref{1RDM_output}) and (\ref{2RDM_output}),
in order to evaluate the whole-molecule 1- and 2-RDMs;\ since LASSCF has a wave function,
this concern is not directly relevant.

Figure \ref{las_1rdm_fragbasis_fig} in Sec.\ \ref{schmidt_analysis} depicts two potential choices of fragments in
a LASSCF calculation: non-overlapping fragments, as in standard DMET, labeled above the 1-RDM,
and overlapping or ``enlarged'' fragments, labeled below the 1-RDM. The latter are attractive because
of the greater freedom they provide in the optimization of the active orbitals, especially because in LASSCF
fewer bath orbitals are available than in standard DMET (see Fig.\ \ref{las_1rdm_embbasis_fig};\ active subspaces
are unentangled and therefore the maximum number of bath orbitals is only $M_{X_K}$, not $M_{F_K}=M_{X_K}+M_{A_K}$).
However, if a future post-LASSCF method sacrifices exact embedding of the impurity and trial wave functions
in order to model dynamical correlation or intersubspace entanglement, then the
correlation potential of standard DMET will have to be invoked and Eqs.\ (\ref{1RDM_output})
and (\ref{2RDM_output}) will have to be used, at which point non-overlapping fragments become mandatory. Therefore,
in our current implementation, we construct both.

Specifically, the ``enlarged fragment orbitals,'' $|\tilde{f}_{Kn}\rangle$ (bottom label of Fig.\ \ref{las_1rdm_fragbasis_fig}), are
a superset of ``fragment orbitals,'' $|f_{Kn}\rangle$ (top label of Fig.\ \ref{las_1rdm_fragbasis_fig}),
\begin{eqnarray}
    \{|\tilde{f}_{Kn}\rangle\} &=& \{|a_{Kn}\rangle\} \cup \{|x_{Kn}\rangle\} \cup \{|\tilde{x}_{Kn}\rangle\},
    \nonumber \\ &=& \{|f_{Kn}\rangle\} \cup \{|\tilde{x}_{Kn}\rangle\},
\end{eqnarray}
where $|x_{Kn}\rangle$ is a non-overlapping external fragment orbital such that all $|f_{Kn}\rangle$
tile the molecule, and $|\tilde{x}_{Kn}\rangle$ is also confined to the external space but overlaps
with external orbitals assigned to other fragments, $\langle f_{Kn}|\tilde{x}_{Ln}\rangle\neq 0$ for $K\neq L$.
For the purposes of Eqs.\ (\ref{single_determinant_schmidt})--(\ref{N_I}), $|\tilde{f}_{Kn}\rangle$ are the fragment orbitals;\ for
the purpose of Eqs.\ (\ref{1RDM_output}) and (\ref{2RDM_output}), $|f_{Kn}\rangle$ are the fragment orbitals.

Both $|x_{Kn}\rangle$ and $|\tilde{x}_{Kn}\rangle$ are related to the
solutions of Eq.\ (\ref{assign_external}) from Sec.\ \ref{update_frag}.
For the non-overlapping $|x_{Kn}\rangle$, we repeatedly cycle over the fragment index $K$, solve Eq.\ (\ref{assign_external}),
and assign up to $M_{X_K}$ solutions with eigenvalue greater than 0.5 to the $K$th fragment.
We then remove those assigned $|x_{Kn}\rangle$ from the projector before continuing to the next fragment,
\begin{eqnarray}
    \hat{Q}_{X_K} \hat{P}_{X_\mathrm{mol}} \hat{Q}_{X_K} \to \hat{P}_{X_\mathrm{mol}},
\end{eqnarray}
where
\begin{eqnarray}
    \hat{Q}_{X_K} &=& 1 - \sum_{x_{Kn}} |x_{Kn}\rangle\langle x_{Kn}|.
\end{eqnarray}
This ensures that $|x_{Kn}\rangle$ for different $K$ are mutually orthogonal. We continue cycling
over $K$ in this manner until all external orbitals are assigned and each fragment has $M_{X_K}$ non-overlapping external fragment orbitals.

The overlapping external fragment orbitals, $|\tilde{x}_{Kn}\rangle$, are constructed by solving
\begin{eqnarray}
    \hat{P}_{\tilde{X}_K}|\tilde{x}_{Kn}\rangle &=& |\tilde{x}_{Kn}\rangle,
    \label{eigenorbs}
\end{eqnarray}
(note unity eigenvalue) where
\begin{eqnarray}
    \hat{P}_{\tilde{X}_K} &=& \sum_{L,M\neq K} \hat{P}_{X_L}\hat{R}_{G_K}\hat{P}_{X_M},
    \label{P_tX_K}
    \\
    \hat{R}_{G_K} &=& \hat{S}_{G_K}^{1/2} \hat{S}_{G_K}^{-1} \hat{S}_{G_K}^{1/2},
    \label{R_G_K}
    \\
    \hat{S}_{G_K} &=& \sum_{L,M} \hat{P}_{X_L} \hat{P}_{G_K} \hat{P}_{X_M}
    \label{S_G_K}
    \\
    \hat{P}_{X_L} &=& \sum_{x_{Ln}} |x_{Ln}\rangle\langle x_{Ln}|.
    \label{P_X_L}
\end{eqnarray}
In words: we project the guess-fragment orbitals, $|g_{Kn}\rangle$,
onto the whole-molecule external space [Eq.\ (\ref{S_G_K})], orthonormalize and remove linear dependencies [Eq.\ (\ref{R_G_K})],
and retain linear combinations of these that are orthogonal to the external orbitals that $K$ already has, $|x_{Kn}\rangle$ [Eqs.\ (\ref{P_tX_K}) and (\ref{eigenorbs})].
This restores degrees of freedom from the guess fragment that were lost by enforcing Eq.\ (\ref{schmidt_cas_condition});\ there are always
as many eigenorbitals of $\hat{P}_{\tilde{X}_K}$ with unity eigenvalue as there are active orbitals, $M_{\tilde{X}_K}=M_{A_K}$, unless some active orbitals
are entirely contained within the guess fragment.

\subsection{CASSCF impurity calculation \label{impsolve_guesses}}

A CASSCF($N_{A_K},M_{A_K}$) calculation for the $K$th impurity is standard, provided the implementation can accept
arbitrary values for the one- and two-electron Hamiltonian integrals. However, the quality of CASSCF 
results depends strongly on the initial guess for the coefficients defining occupied, active, and virtual MOs which is provided to the solver.
In both LASSCF and CAS-DMET, Schmidt decomposition complicates this guess by rotating both occupied and virtual orbitals
among themselves.

In the first iteration of both LASSCF and DMET, a set of guesses for $|a_{Kn}\rangle$ need to be explicitly constructed. Let $|a_n\rangle$
be a reasonable (unlocalized) guess for an active orbital of a comparable CASSCF($N_A$, $M_A$) calculation,
where $N_A=\sum_K N_{A_K}$ and $M_A=\sum_K M_{A_K}$. We generate reasonable guesses for $|a_{Kn}\rangle$ by projecting
the guess-active orbitals onto the guess-fragment space:
\begin{eqnarray}
    \hat{P}_{G_K}\hat{P}_{A_\mathrm{mol}}|a_{Kn}\rangle &=& \lambda_{a_{Kn}}|a_{Kn}\rangle,
    \label{guess_active_orbs}
\end{eqnarray}
retaining the $M_{A_K}$ highest-eigenvalue solutions, where
\begin{eqnarray}
    \hat{P}_{A_\mathrm{mol}} &\equiv& \sum_{a_n}^{M_A} |a_n\rangle\langle a_n|.
\end{eqnarray}
Of course, we also import (and project) converged $|a_{Kn}\rangle$ from adjacent points on the potential energy surface, if we are
doing a potential energy scan.

In general, a set of guesses for $|a_{Kn}\rangle$ may not be contained
in the impurity space, since bath orbitals cannot be perfectly anticipated ahead of time.
After Schmidt decomposition, the active-orbital guesses must be projected into the impurity space,
\begin{eqnarray}
    \hat{P}_{I_K}\hat{P}_{A_K}|p_{Kn}\rangle &=& \lambda_{p_{Kn}}|p_{Kn}\rangle,
    \label{guess_mo_eigenorb1}
\end{eqnarray}
where
\begin{eqnarray}
    \hat{P}_{A_K} &=& \sum_{a_{Kn}}^{M_{A_K}} |a_{Kn}\rangle\langle a_{Kn}|,
    \label{P_A_K}
    \\
    \hat{P}_{I_K} &=& \sum_{i_{Kn}}^{M_{I_K}} |i_{Kn}\rangle\langle i_{Kn}|.
    \label{P_I_K_orb}
\end{eqnarray}
[$\hat{P}_{I_K}$ is here a one-body operator, as opposed to the many-body projector of Eq.\ (\ref{P_I_K}).]
We retain all of the solutions to Eq.\ (\ref{guess_mo_eigenorb1}). Sorting them from smallest eigenvalue to largest,
we then also solve
\begin{eqnarray}
    \hat{Q}_{A_K}\hat{h}_{I_K}\hat{Q}_{A_K}|p_{Kn}\rangle &=& \epsilon_{p_{Kn}}|p_{Kn}\rangle,
    \label{guess_mo_eigenorb2}
\end{eqnarray}
and again sort solutions from lowest to highest eigenvalue,
where $\hat{h}_{I_K}$ is the one-body Hamiltonian or a realistic Fock operator for the $K$th impurity,
and where
\begin{eqnarray}
    \hat{Q}_{A_K} &\equiv& 1-\hat{P}_{A_K} = \sum_{p_{Kn}}^{M_{I_K}-M_{A_K}} |p_{Kn}\rangle\langle p_{Kn}|.
    \label{Q_A_K}
\end{eqnarray}
Thus, of the complete sorted set of $|p_{Kn}\rangle$, the first $N_{I_K}-N_{A_K}$ (in terms of spinorbitals)
are the guesses for the occupied MOs, the last $M_{A_K}$ are
the guesses for the active MOs, and the remainder are the guesses for the virtual MOs.

\begin{tocentry}
\includegraphics{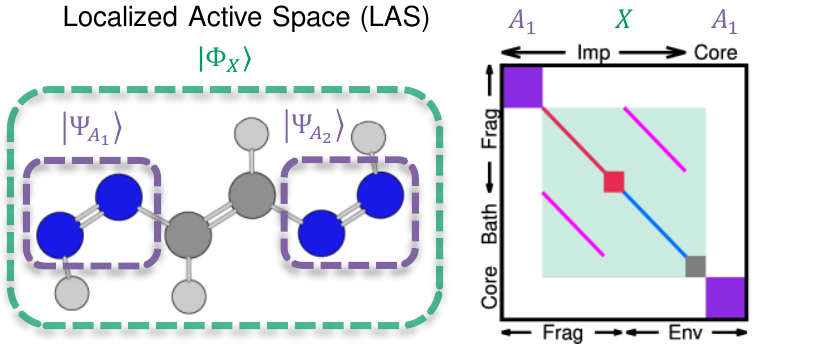}
\end{tocentry}

\bibliography{lib.bib}

\providecommand{\latin}[1]{#1}
\makeatletter
\providecommand{\doi}
  {\begingroup\let\do\@makeother\dospecials
  \catcode`\{=1 \catcode`\}=2\doi@aux}
\providecommand{\doi@aux}[1]{\endgroup\texttt{#1}}
\makeatother
\providecommand*\mcitethebibliography{\thebibliography}
\csname @ifundefined\endcsname{endmcitethebibliography}
  {\let\endmcitethebibliography\endthebibliography}{}
\begin{mcitethebibliography}{53}
\providecommand*\natexlab[1]{#1}
\providecommand*\mciteSetBstSublistMode[1]{}
\providecommand*\mciteSetBstMaxWidthForm[2]{}
\providecommand*\mciteBstWouldAddEndPuncttrue
  {\def\EndOfBibitem{\unskip.}}
\providecommand*\mciteBstWouldAddEndPunctfalse
  {\let\EndOfBibitem\relax}
\providecommand*\mciteSetBstMidEndSepPunct[3]{}
\providecommand*\mciteSetBstSublistLabelBeginEnd[3]{}
\providecommand*\EndOfBibitem{}
\mciteSetBstSublistMode{f}
\mciteSetBstMaxWidthForm{subitem}{(\alph{mcitesubitemcount})}
\mciteSetBstSublistLabelBeginEnd
  {\mcitemaxwidthsubitemform\space}
  {\relax}
  {\relax}

\bibitem[Jacob and Neugebauer(2014)Jacob, and Neugebauer]{Jacob2014}
Jacob,~C.~R.; Neugebauer,~J. {Subsystem density-functional theory}. \emph{WIREs
  Comput. Mol. Sci.} \textbf{2014}, \emph{4}, 325\relax
\mciteBstWouldAddEndPuncttrue
\mciteSetBstMidEndSepPunct{\mcitedefaultmidpunct}
{\mcitedefaultendpunct}{\mcitedefaultseppunct}\relax
\EndOfBibitem
\bibitem[Wesolowski \latin{et~al.}(2015)Wesolowski, Shedge, and
  Zhou]{Wesolowski2015}
Wesolowski,~T.~A.; Shedge,~S.; Zhou,~X. {Frozen-Density Embedding Strategy for
  Multilevel Simulations of Electronic Structure}. \emph{Chem. Rev.}
  \textbf{2015}, \emph{115}, 5891\relax
\mciteBstWouldAddEndPuncttrue
\mciteSetBstMidEndSepPunct{\mcitedefaultmidpunct}
{\mcitedefaultendpunct}{\mcitedefaultseppunct}\relax
\EndOfBibitem
\bibitem[Collins and Bettens(2015)Collins, and Bettens]{Collins2015}
Collins,~M.~A.; Bettens,~R. P.~A. {Energy-Based Molecular Fragmentation
  Methods}. \emph{Chem. Rev.} \textbf{2015}, \emph{115}, 5607\relax
\mciteBstWouldAddEndPuncttrue
\mciteSetBstMidEndSepPunct{\mcitedefaultmidpunct}
{\mcitedefaultendpunct}{\mcitedefaultseppunct}\relax
\EndOfBibitem
\bibitem[Raghavachari and Saha(2015)Raghavachari, and Saha]{Raghavachari2015}
Raghavachari,~K.; Saha,~A. {Accurate Composite and Fragment-Based Quantum
  Chemical Models for Large Molecules}. \emph{Chem. Rev.} \textbf{2015},
  \emph{115}, 5643\relax
\mciteBstWouldAddEndPuncttrue
\mciteSetBstMidEndSepPunct{\mcitedefaultmidpunct}
{\mcitedefaultendpunct}{\mcitedefaultseppunct}\relax
\EndOfBibitem
\bibitem[Sun and Chan(2016)Sun, and Chan]{Sun2016}
Sun,~Q.; Chan,~G. K.-L. {Quantum Embedding Theories}. \emph{Acc. Chem. Res}
  \textbf{2016}, \emph{49}, 2705\relax
\mciteBstWouldAddEndPuncttrue
\mciteSetBstMidEndSepPunct{\mcitedefaultmidpunct}
{\mcitedefaultendpunct}{\mcitedefaultseppunct}\relax
\EndOfBibitem
\bibitem[Li \latin{et~al.}(2007)Li, Li, and Jiang]{Li2007}
Li,~W.; Li,~S.; Jiang,~Y. {Generalized Energy-Based Fragmentation Approach for
  Computing the Ground-State Energies and Properties of Large Molecules}.
  \emph{J. Phys. Chem. A} \textbf{2007}, \emph{111}, 2193\relax
\mciteBstWouldAddEndPuncttrue
\mciteSetBstMidEndSepPunct{\mcitedefaultmidpunct}
{\mcitedefaultendpunct}{\mcitedefaultseppunct}\relax
\EndOfBibitem
\bibitem[Isegawa \latin{et~al.}(2013)Isegawa, Wang, and Truhlar]{Isegawa2013}
Isegawa,~M.; Wang,~B.; Truhlar,~D.~G. {Electrostatically Embedded Molecular
  Tailoring Approach and Validation for Peptides}. \emph{J. Chem. Theory
  Comput.} \textbf{2013}, \emph{9}, 1381\relax
\mciteBstWouldAddEndPuncttrue
\mciteSetBstMidEndSepPunct{\mcitedefaultmidpunct}
{\mcitedefaultendpunct}{\mcitedefaultseppunct}\relax
\EndOfBibitem
\bibitem[Collins \latin{et~al.}(2014)Collins, Cvitkovic, and
  Bettens]{Collins2014}
Collins,~M.~A.; Cvitkovic,~M.~W.; Bettens,~R. P.~A. {The Combined Fragmentation
  and Systematic Molecular Fragmentation Methods}. \emph{Acc. Chem. Res.}
  \textbf{2014}, \emph{47}, 2776\relax
\mciteBstWouldAddEndPuncttrue
\mciteSetBstMidEndSepPunct{\mcitedefaultmidpunct}
{\mcitedefaultendpunct}{\mcitedefaultseppunct}\relax
\EndOfBibitem
\bibitem[Kamiya \latin{et~al.}(2008)Kamiya, Hirata, and Valiev]{Kamiya2008}
Kamiya,~M.; Hirata,~S.; Valiev,~M. {Fast electron correlation methods for
  molecular clusters without basis set superposition errors.} \emph{The Journal
  of chemical physics} \textbf{2008}, \emph{128}, 074103\relax
\mciteBstWouldAddEndPuncttrue
\mciteSetBstMidEndSepPunct{\mcitedefaultmidpunct}
{\mcitedefaultendpunct}{\mcitedefaultseppunct}\relax
\EndOfBibitem
\bibitem[Richard and Herbert(2012)Richard, and Herbert]{Richard2012}
Richard,~R.~M.; Herbert,~J.~M. {A generalized many-body expansion and a unified
  view of fragment-based methods in electronic structure theory}. \emph{J.
  Chem. Phys.} \textbf{2012}, \emph{137}, 064113\relax
\mciteBstWouldAddEndPuncttrue
\mciteSetBstMidEndSepPunct{\mcitedefaultmidpunct}
{\mcitedefaultendpunct}{\mcitedefaultseppunct}\relax
\EndOfBibitem
\bibitem[Cortona(1991)]{Cortona1991}
Cortona,~P. {Self-consistently determined properties of solids without
  band-structure calculations}. \emph{Phys. Rev. B} \textbf{1991}, \emph{44},
  8454\relax
\mciteBstWouldAddEndPuncttrue
\mciteSetBstMidEndSepPunct{\mcitedefaultmidpunct}
{\mcitedefaultendpunct}{\mcitedefaultseppunct}\relax
\EndOfBibitem
\bibitem[Wesolowskv and Warshel(1993)Wesolowskv, and Warshel]{Wesolowskv1993}
Wesolowskv,~T.~A.; Warshel,~A. {Frozen Density Functional Approach for ab
  Initio Calculations of Solvated Molecules}. \emph{J. Phys. Chem}
  \textbf{1993}, \emph{97}, 8050--8053\relax
\mciteBstWouldAddEndPuncttrue
\mciteSetBstMidEndSepPunct{\mcitedefaultmidpunct}
{\mcitedefaultendpunct}{\mcitedefaultseppunct}\relax
\EndOfBibitem
\bibitem[Georges \latin{et~al.}(1996)Georges, Kotliar, Krauth, and
  Rozenberg]{Georges1996}
Georges,~A.; Kotliar,~G.; Krauth,~W.; Rozenberg,~M.~J. {Dynamical mean-field
  theory of strongly correlated fermion systems and the limit of infinite
  dimensions}. \emph{Rev. Mod. Phys.} \textbf{1996}, \emph{68}, 13\relax
\mciteBstWouldAddEndPuncttrue
\mciteSetBstMidEndSepPunct{\mcitedefaultmidpunct}
{\mcitedefaultendpunct}{\mcitedefaultseppunct}\relax
\EndOfBibitem
\bibitem[Kananenka \latin{et~al.}(2015)Kananenka, Gull, and
  Zgid]{Kananenka2015}
Kananenka,~A.~A.; Gull,~E.; Zgid,~D. {Systematically improvable multiscale
  solver for correlated electron systems}. \emph{Phys. Rev. B} \textbf{2015},
  \emph{91}, 121111(R)\relax
\mciteBstWouldAddEndPuncttrue
\mciteSetBstMidEndSepPunct{\mcitedefaultmidpunct}
{\mcitedefaultendpunct}{\mcitedefaultseppunct}\relax
\EndOfBibitem
\bibitem[Knizia and Chan(2012)Knizia, and Chan]{Knizia2012}
Knizia,~G.; Chan,~G. K.-l. {Density Matrix Embedding : A Simple Alternative to
  Dynamical Mean-Field Theory}. \emph{Phys. Rev. Lett.} \textbf{2012},
  \emph{109}, 186404\relax
\mciteBstWouldAddEndPuncttrue
\mciteSetBstMidEndSepPunct{\mcitedefaultmidpunct}
{\mcitedefaultendpunct}{\mcitedefaultseppunct}\relax
\EndOfBibitem
\bibitem[Wouters \latin{et~al.}(2016)Wouters, Jim{\'{e}}nez-Hoyos, Sun, and
  Chan]{Wouters2016}
Wouters,~S.; Jim{\'{e}}nez-Hoyos,~C.~A.; Sun,~Q.; Chan,~G. K.~L. {A Practical
  Guide to Density Matrix Embedding Theory in Quantum Chemistry}. \emph{J.
  Chem. Theory Comput} \textbf{2016}, \emph{12}, 2706--2719\relax
\mciteBstWouldAddEndPuncttrue
\mciteSetBstMidEndSepPunct{\mcitedefaultmidpunct}
{\mcitedefaultendpunct}{\mcitedefaultseppunct}\relax
\EndOfBibitem
\bibitem[Wouters \latin{et~al.}(2017)Wouters, Jim{\'{e}}nez-Hoyos, and
  Chan]{Wouters2017}
Wouters,~S.; Jim{\'{e}}nez-Hoyos,~C.~A.; Chan,~G. K.~L. In \emph{Fragmentation:
  Toward Accurate Calculations on Complex Molecular Systems}; Gordon,~M.~S.,
  Ed.; Wiley, 2017; Chapter 8, p 227\relax
\mciteBstWouldAddEndPuncttrue
\mciteSetBstMidEndSepPunct{\mcitedefaultmidpunct}
{\mcitedefaultendpunct}{\mcitedefaultseppunct}\relax
\EndOfBibitem
\bibitem[Roos(2005)]{ROOS2005725}
Roos,~B.~O. In \emph{Theory and Applications of Computational Chemistry};
  Dykstra,~C.~E., Frenking,~G., Kim,~K.~S., Scuseria,~G.~E., Eds.; Elsevier:
  Amsterdam, 2005; pp 725--764\relax
\mciteBstWouldAddEndPuncttrue
\mciteSetBstMidEndSepPunct{\mcitedefaultmidpunct}
{\mcitedefaultendpunct}{\mcitedefaultseppunct}\relax
\EndOfBibitem
\bibitem[Schmidt and Gordon(1998)Schmidt, and Gordon]{Schmidt1998}
Schmidt,~M.~W.; Gordon,~M.~S. {The Construction and Interpretation of MCSCF
  Wavefunctions}. \emph{Annu. Rev. Phys. Chem} \textbf{1998}, \emph{49},
  233\relax
\mciteBstWouldAddEndPuncttrue
\mciteSetBstMidEndSepPunct{\mcitedefaultmidpunct}
{\mcitedefaultendpunct}{\mcitedefaultseppunct}\relax
\EndOfBibitem
\bibitem[Roos \latin{et~al.}(1980)Roos, Taylor, and Siegbahn]{Roos1980}
Roos,~B.~O.; Taylor,~P.~R.; Siegbahn,~P. E.~M. {A Complete Active Space SCF
  Method (CASSCF) Using a Density Matrix Formulated Super-CI Approach}.
  \emph{Chem. Phys.} \textbf{1980}, \emph{48}, 157\relax
\mciteBstWouldAddEndPuncttrue
\mciteSetBstMidEndSepPunct{\mcitedefaultmidpunct}
{\mcitedefaultendpunct}{\mcitedefaultseppunct}\relax
\EndOfBibitem
\bibitem[Olsen \latin{et~al.}(1988)Olsen, Roos, J{\o}rgensen, and
  Jensen]{Olsen1988}
Olsen,~J.; Roos,~B.~O.; J{\o}rgensen,~P.; Jensen,~H. J.~A. {Determinant based
  configuration interaction algorithms for complete and restricted
  configuration interaction spaces}. \emph{J. Chem. Phys.} \textbf{1988},
  \emph{89}, 2185\relax
\mciteBstWouldAddEndPuncttrue
\mciteSetBstMidEndSepPunct{\mcitedefaultmidpunct}
{\mcitedefaultendpunct}{\mcitedefaultseppunct}\relax
\EndOfBibitem
\bibitem[Malmqvist \latin{et~al.}(1990)Malmqvist, Rendell, and
  Roos]{Malmqvist1990}
Malmqvist,~P.-{\'{A}}.; Rendell,~A.; Roos,~B.~O. {The Restricted Active Space
  Self-Consistent-Fleld Method, Implemented with a Split Graph Unitary Group
  Approach}. \emph{J. Phys. Chem.} \textbf{1990}, \emph{94}, 5477\relax
\mciteBstWouldAddEndPuncttrue
\mciteSetBstMidEndSepPunct{\mcitedefaultmidpunct}
{\mcitedefaultendpunct}{\mcitedefaultseppunct}\relax
\EndOfBibitem
\bibitem[Ma \latin{et~al.}(2011)Ma, Manni, and Gagliardi]{Ma2011}
Ma,~D.; Manni,~G.~L.; Gagliardi,~L. {The generalized active space concept in
  multiconfigurational self-consistent field methods}. \emph{J. Chem. Phys.}
  \textbf{2011}, \emph{135}, 044128\relax
\mciteBstWouldAddEndPuncttrue
\mciteSetBstMidEndSepPunct{\mcitedefaultmidpunct}
{\mcitedefaultendpunct}{\mcitedefaultseppunct}\relax
\EndOfBibitem
\bibitem[Ivanic(2003)]{Ivanic2003}
Ivanic,~J. {Direct configuration interaction and multiconfigurational
  self-consistent-field method for multiple active spaces with variable
  occupations. I. Method}. \emph{J. Chem. Phys.} \textbf{2003}, \emph{119},
  9364\relax
\mciteBstWouldAddEndPuncttrue
\mciteSetBstMidEndSepPunct{\mcitedefaultmidpunct}
{\mcitedefaultendpunct}{\mcitedefaultseppunct}\relax
\EndOfBibitem
\bibitem[Horike \latin{et~al.}(2009)Horike, Shimomura, and
  Kitagawa]{Horike2009}
Horike,~S.; Shimomura,~S.; Kitagawa,~S. {Soft porous crystals}. \emph{Nat.
  Chem.} \textbf{2009}, \emph{1}, 695--704\relax
\mciteBstWouldAddEndPuncttrue
\mciteSetBstMidEndSepPunct{\mcitedefaultmidpunct}
{\mcitedefaultendpunct}{\mcitedefaultseppunct}\relax
\EndOfBibitem
\bibitem[Lee \latin{et~al.}(2009)Lee, Farha, Roberts, Scheidt, Nguyen, and
  Hupp]{Lee2009a}
Lee,~J.; Farha,~O.~K.; Roberts,~J.; Scheidt,~K.~A.; Nguyen,~S.~T.; Hupp,~J.~T.
  {Metal-organic framework materials as catalysts}. \emph{Chem. Soc. Rev.}
  \textbf{2009}, \emph{38}, 1450--1459\relax
\mciteBstWouldAddEndPuncttrue
\mciteSetBstMidEndSepPunct{\mcitedefaultmidpunct}
{\mcitedefaultendpunct}{\mcitedefaultseppunct}\relax
\EndOfBibitem
\bibitem[Odoh \latin{et~al.}(2015)Odoh, Cramer, Truhlar, and
  Gagliardi]{Odoh2015}
Odoh,~S.~O.; Cramer,~C.~J.; Truhlar,~D.~G.; Gagliardi,~L. {Quantum-Chemical
  Characterization of the Properties and Reactivities of Metal−Organic
  Frameworks}. \emph{Chem. Rev.} \textbf{2015}, \emph{115}, 6051\relax
\mciteBstWouldAddEndPuncttrue
\mciteSetBstMidEndSepPunct{\mcitedefaultmidpunct}
{\mcitedefaultendpunct}{\mcitedefaultseppunct}\relax
\EndOfBibitem
\bibitem[Coudert and Fuchs(2016)Coudert, and Fuchs]{Coudert2016}
Coudert,~F.-X.; Fuchs,~A.~H. {Computational characterization and prediction of
  metal–organic framework properties}. \emph{Coord. Chem. Rev.}
  \textbf{2016}, \emph{307}, 211\relax
\mciteBstWouldAddEndPuncttrue
\mciteSetBstMidEndSepPunct{\mcitedefaultmidpunct}
{\mcitedefaultendpunct}{\mcitedefaultseppunct}\relax
\EndOfBibitem
\bibitem[Yan(2017)]{Yan2017}
Yan,~B. {Lanthanide-Functionalized Metal−Organic Framework Hybrid Systems To
  Create Multiple Luminescent Centers for Chemical Sensing}. \emph{Acc. Chem.
  Res.} \textbf{2017}, \emph{50}, 2789\relax
\mciteBstWouldAddEndPuncttrue
\mciteSetBstMidEndSepPunct{\mcitedefaultmidpunct}
{\mcitedefaultendpunct}{\mcitedefaultseppunct}\relax
\EndOfBibitem
\bibitem[Bernales \latin{et~al.}(2018)Bernales, Ortu{\~{n}}o, Truhlar, Cramer,
  and Gagliardi]{Bernales2018}
Bernales,~V.; Ortu{\~{n}}o,~M.~A.; Truhlar,~D.~G.; Cramer,~C.~J.; Gagliardi,~L.
  {Computational Design of Functionalized Metal − Organic Framework Nodes for
  Catalysis}. \emph{ACS Cent. Sci.} \textbf{2018}, \emph{4}, 5--19\relax
\mciteBstWouldAddEndPuncttrue
\mciteSetBstMidEndSepPunct{\mcitedefaultmidpunct}
{\mcitedefaultendpunct}{\mcitedefaultseppunct}\relax
\EndOfBibitem
\bibitem[Schmidt(1907)]{Schmidt1907}
Schmidt,~E. {Zur Theorie der Linearen und Nichtlinearen Integralgleichungen. I
  Teil. Entwicklung Willk{\"{u}}rlichen Funktionen nach System
  Vorgeschriebener}. \emph{Math. Annalen} \textbf{1907}, \emph{63}, 433\relax
\mciteBstWouldAddEndPuncttrue
\mciteSetBstMidEndSepPunct{\mcitedefaultmidpunct}
{\mcitedefaultendpunct}{\mcitedefaultseppunct}\relax
\EndOfBibitem
\bibitem[Peschel(2012)]{Peschel2012}
Peschel,~I. {Special Review: Entanglement in Solvable Many-Particle Models}.
  \emph{Braz J Phys} \textbf{2012}, \emph{42}, 267--291\relax
\mciteBstWouldAddEndPuncttrue
\mciteSetBstMidEndSepPunct{\mcitedefaultmidpunct}
{\mcitedefaultendpunct}{\mcitedefaultseppunct}\relax
\EndOfBibitem
\bibitem[Waecziel and Levitt(1976)Waecziel, and Levitt]{Waecziel1976}
Waecziel,~A.; Levitt,~M. {Theoretical Studies of Enzymic Reactions :
  Dielectric, Electrostatic and Steric Stabilization of the Carbonium Ion in
  the Reaction of Lysozyme}. \emph{J. Mol. Biol} \textbf{1976}, \emph{103},
  227--249\relax
\mciteBstWouldAddEndPuncttrue
\mciteSetBstMidEndSepPunct{\mcitedefaultmidpunct}
{\mcitedefaultendpunct}{\mcitedefaultseppunct}\relax
\EndOfBibitem
\bibitem[Lin and Truhlar(2007)Lin, and Truhlar]{Lin2007}
Lin,~H.; Truhlar,~D.~G. {QM/MM: what have we learned, where are we, and where
  do we go from here?} \emph{Theor. Chem. Acc.} \textbf{2007}, \emph{117},
  185--199\relax
\mciteBstWouldAddEndPuncttrue
\mciteSetBstMidEndSepPunct{\mcitedefaultmidpunct}
{\mcitedefaultendpunct}{\mcitedefaultseppunct}\relax
\EndOfBibitem
\bibitem[Knizia and Chan(2013)Knizia, and Chan]{Knizia2013}
Knizia,~G.; Chan,~G. K.-l. {Density Matrix Embedding: A Strong-Coupling Quantum
  Embedding Theory}. \emph{J. Chem. Theory Comput.} \textbf{2013}, 1428\relax
\mciteBstWouldAddEndPuncttrue
\mciteSetBstMidEndSepPunct{\mcitedefaultmidpunct}
{\mcitedefaultendpunct}{\mcitedefaultseppunct}\relax
\EndOfBibitem
\bibitem[Pham \latin{et~al.}(2018)Pham, Bernales, and Gagliardi]{Pham2018}
Pham,~H.~Q.; Bernales,~V.; Gagliardi,~L. {Can Density Matrix Embedding Theory
  with the Complete Activate Space Self-Consistent Field Solver Describe Single
  and Double Bond Breaking in Molecular Systems?} \emph{J. Chem. Theory
  Comput.} \textbf{2018}, \emph{14}, 1960\relax
\mciteBstWouldAddEndPuncttrue
\mciteSetBstMidEndSepPunct{\mcitedefaultmidpunct}
{\mcitedefaultendpunct}{\mcitedefaultseppunct}\relax
\EndOfBibitem
\bibitem[Sandhoefer and Chan(2016)Sandhoefer, and Chan]{Sandhoefer2016}
Sandhoefer,~B.; Chan,~G. K.-L. {Density matrix embedding theory for interacting
  electron-phonon systems}. \emph{Phys. Rev. B} \textbf{2016}, \emph{94},
  085115\relax
\mciteBstWouldAddEndPuncttrue
\mciteSetBstMidEndSepPunct{\mcitedefaultmidpunct}
{\mcitedefaultendpunct}{\mcitedefaultseppunct}\relax
\EndOfBibitem
\bibitem[Tsuchimochi \latin{et~al.}(2015)Tsuchimochi, Welborn, Voorhis, and
  {Van Voorhis}]{Tsuchimochi2015}
Tsuchimochi,~T.; Welborn,~M.; Voorhis,~T.~V.; {Van Voorhis},~T. {Density matrix
  embedding in an antisymmetrized geminal power bath}. \emph{J. Chem. Phys.}
  \textbf{2015}, \emph{143}, 024107\relax
\mciteBstWouldAddEndPuncttrue
\mciteSetBstMidEndSepPunct{\mcitedefaultmidpunct}
{\mcitedefaultendpunct}{\mcitedefaultseppunct}\relax
\EndOfBibitem
\bibitem[Fan and Jie(2015)Fan, and Jie]{Fan2015}
Fan,~Z.; Jie,~Q.-l. {Cluster density matrix embedding theory for quantum spin
  systems}. \emph{Phys. Rev. B} \textbf{2015}, \emph{91}, 195118\relax
\mciteBstWouldAddEndPuncttrue
\mciteSetBstMidEndSepPunct{\mcitedefaultmidpunct}
{\mcitedefaultendpunct}{\mcitedefaultseppunct}\relax
\EndOfBibitem
\bibitem[Gunst \latin{et~al.}(2017)Gunst, Wouters, {De Baerdemacker}, and {Van
  Neck}]{Gunst2017}
Gunst,~K.; Wouters,~S.; {De Baerdemacker},~S.; {Van Neck},~D. {Block product
  density matrix embedding theory for strongly correlated spin systems}.
  \emph{Phys. Rev. B} \textbf{2017}, \emph{95}, 195127\relax
\mciteBstWouldAddEndPuncttrue
\mciteSetBstMidEndSepPunct{\mcitedefaultmidpunct}
{\mcitedefaultendpunct}{\mcitedefaultseppunct}\relax
\EndOfBibitem
\bibitem[March \latin{et~al.}(1995)March, Young, and
  Sampanthar]{MarchYoungSampanthar_mbpt}
March,~N.~H.; Young,~W.~H.; Sampanthar,~S. \emph{The Many-Body Problem in
  Quantum Mechanics}; Dover Publications, Inc.: Mineola, N.Y., 1995; Chapter 4,
  pp 67--135\relax
\mciteBstWouldAddEndPuncttrue
\mciteSetBstMidEndSepPunct{\mcitedefaultmidpunct}
{\mcitedefaultendpunct}{\mcitedefaultseppunct}\relax
\EndOfBibitem
\bibitem[Shavitt and Bartlett(2009)Shavitt, and Bartlett]{ShavittBartlett_mbpt}
Shavitt,~I.; Bartlett,~R.~J. \emph{Many-Body Methods in Chemistry and Physics};
  Cambridge University Press: New York, 2009; Chapter 5, pp 130--164\relax
\mciteBstWouldAddEndPuncttrue
\mciteSetBstMidEndSepPunct{\mcitedefaultmidpunct}
{\mcitedefaultendpunct}{\mcitedefaultseppunct}\relax
\EndOfBibitem
\bibitem[Foster and Boys(1960)Foster, and Boys]{Foster1960}
Foster,~J.~M.; Boys,~S.~F. {Canonical Configurational Interaction Procedure}.
  \emph{Rev. Mod. Phys.} \textbf{1960}, \emph{32}, 300\relax
\mciteBstWouldAddEndPuncttrue
\mciteSetBstMidEndSepPunct{\mcitedefaultmidpunct}
{\mcitedefaultendpunct}{\mcitedefaultseppunct}\relax
\EndOfBibitem
\bibitem[Sun and Chan(2014)Sun, and Chan]{Sun2014}
Sun,~Q.; Chan,~G. K.~L. {Exact and optimal quantum mechanics/molecular
  mechanics boundaries}. \emph{J. Chem. Theory Comput.} \textbf{2014},
  \emph{10}, 3784--3790\relax
\mciteBstWouldAddEndPuncttrue
\mciteSetBstMidEndSepPunct{\mcitedefaultmidpunct}
{\mcitedefaultendpunct}{\mcitedefaultseppunct}\relax
\EndOfBibitem
\bibitem[Sun \latin{et~al.}(2018)Sun, Berkelbach, Blunt, Booth, Guo, Li, Liu,
  McClain, Sayfutyarova, Sharma, Wouters, and Chan]{Sun2018}
Sun,~Q.; Berkelbach,~T.~C.; Blunt,~N.~S.; Booth,~G.~H.; Guo,~S.; Li,~Z.;
  Liu,~J.; McClain,~J.~D.; Sayfutyarova,~E.~R.; Sharma,~S.; Wouters,~S.;
  Chan,~G. K.~L. {PySCF: the Python-based simulations of chemistry framework}.
  \emph{WIREs Comput Mol Sci} \textbf{2018}, \emph{8}, e1340\relax
\mciteBstWouldAddEndPuncttrue
\mciteSetBstMidEndSepPunct{\mcitedefaultmidpunct}
{\mcitedefaultendpunct}{\mcitedefaultseppunct}\relax
\EndOfBibitem
\bibitem[Kutzelnigg and Mukherjee(1999)Kutzelnigg, and
  Mukherjee]{Kutzelnigg1999}
Kutzelnigg,~W.; Mukherjee,~D. {Cumulant expansion of the reduced density
  matrices}. \emph{J. Chem. Phys.} \textbf{1999}, \emph{110}, 2800\relax
\mciteBstWouldAddEndPuncttrue
\mciteSetBstMidEndSepPunct{\mcitedefaultmidpunct}
{\mcitedefaultendpunct}{\mcitedefaultseppunct}\relax
\EndOfBibitem
\bibitem[Frisch \latin{et~al.}(2013)Frisch, Trucks, Schlegel, Scuseria, Robb,
  Cheeseman, Scalmani, Barone, Mennucci, Petersson, Nakatsuji, Caricato, Li,
  Hratchian, Izmaylov, Bloino, Zheng, Sonnenberg, Hada, Ehara, Toyota, Fukuda,
  Hasegawa, Ishida, Nakajima, Honda, Kitao, Nakai, Vreven, Montgomery, Jr.,
  Peralta, Ogliaro, Bearpark, Heyd, Brothers, Kudin, Staroverov, Keith,
  Kobayashi, Normand, Raghavachari, Rendell, Burant, Iyengar, Tomasi, Cossi,
  Rega, Millam, Klene, Knox, Cross, Bakken, Adamo, Jaramillo, Gomperts,
  Stratmann, Yazyev, Austin, Cammi, Pomelli, Ochterski, Martin, Morokuma,
  Zakrzewski, Voth, Salvador, Dannenberg, Dapprich, Daniels, Farkas, Foresman,
  Ortiz, Cioslowski, and Fox]{g09_e01}
Frisch,~M.~J.; Trucks,~G.~W.; Schlegel,~H.~B.; Scuseria,~G.~E.; Robb,~M.~A.;
  Cheeseman,~J.~R.; Scalmani,~G.; Barone,~V.; Mennucci,~B.; Petersson,~G.~A.;
  Nakatsuji,~H.; Caricato,~M.; Li,~X.; Hratchian,~H.~P.; Izmaylov,~A.~F.;
  Bloino,~J.; Zheng,~G.; Sonnenberg,~J.~L.; Hada,~M.; Ehara,~M.; Toyota,~K.;
  Fukuda,~R.; Hasegawa,~J.; Ishida,~M.; Nakajima,~T.; Honda,~Y.; Kitao,~O.;
  Nakai,~H.; Vreven,~T.; Montgomery,~J.~A.; Jr.,; Peralta,~J.~E.; Ogliaro,~F.;
  Bearpark,~M.; Heyd,~J.~J.; Brothers,~E.; Kudin,~K.~N.; Staroverov,~V.~N.;
  Keith,~T.; Kobayashi,~R.; Normand,~J.; Raghavachari,~K.; Rendell,~A.;
  Burant,~J.~C.; Iyengar,~S.~S.; Tomasi,~J.; Cossi,~M.; Rega,~N.;
  Millam,~J.~M.; Klene,~M.; Knox,~J.~E.; Cross,~J.~B.; Bakken,~V.; Adamo,~C.;
  Jaramillo,~J.; Gomperts,~R.; Stratmann,~R.~E.; Yazyev,~O.; Austin,~A.~J.;
  Cammi,~R.; Pomelli,~C.; Ochterski,~J.~W.; Martin,~R.~L.; Morokuma,~K.;
  Zakrzewski,~V.~G.; Voth,~G.~A.; Salvador,~P.; Dannenberg,~J.~J.;
  Dapprich,~S.; Daniels,~A.~D.; Farkas,~O.; Foresman,~J.~B.; Ortiz,~J.~V.;
  Cioslowski,~J.; Fox,~D.~J. {Gaussian 09 Revision E.01}. 2013\relax
\mciteBstWouldAddEndPuncttrue
\mciteSetBstMidEndSepPunct{\mcitedefaultmidpunct}
{\mcitedefaultendpunct}{\mcitedefaultseppunct}\relax
\EndOfBibitem
\bibitem[Hermes(2018)]{mrh_software}
Hermes,~M.~R. {https://github.com/MatthewRHermes/mrh}. 2018;
  \url{https://github.com/MatthewRHermes/mrh}\relax
\mciteBstWouldAddEndPuncttrue
\mciteSetBstMidEndSepPunct{\mcitedefaultmidpunct}
{\mcitedefaultendpunct}{\mcitedefaultseppunct}\relax
\EndOfBibitem
\bibitem[Pham(2018)]{pham_software}
Pham,~H.~Q. https://github.com/hungpham2017/casdmet. 2018;
  \url{https://github.com/hungpham2017/casdmet}\relax
\mciteBstWouldAddEndPuncttrue
\mciteSetBstMidEndSepPunct{\mcitedefaultmidpunct}
{\mcitedefaultendpunct}{\mcitedefaultseppunct}\relax
\EndOfBibitem
\bibitem[Wouters(2016)]{qcdmet_software}
Wouters,~S. https://github.com/sebwouters/qc-dmet. 2016;
  \url{https://github.com/sebwouters/qc-dmet}\relax
\mciteBstWouldAddEndPuncttrue
\mciteSetBstMidEndSepPunct{\mcitedefaultmidpunct}
{\mcitedefaultendpunct}{\mcitedefaultseppunct}\relax
\EndOfBibitem
\bibitem[Andersson \latin{et~al.}(1992)Andersson, Malmqvist, and
  Roos]{Andersson1992}
Andersson,~K.; Malmqvist,~P.-A.; Roos,~B.~O. {Second-order perturbation theory
  with a complete active space self-consistent field reference function}.
  \emph{J. Chem. Phys.} \textbf{1992}, \emph{96}, 1218\relax
\mciteBstWouldAddEndPuncttrue
\mciteSetBstMidEndSepPunct{\mcitedefaultmidpunct}
{\mcitedefaultendpunct}{\mcitedefaultseppunct}\relax
\EndOfBibitem
\bibitem[Manni \latin{et~al.}(2014)Manni, Carlson, Luo, Ma, Olsen, Truhlar, and
  Gagliardi]{Manni2014}
Manni,~G.~L.; Carlson,~R.~K.; Luo,~S.; Ma,~D.; Olsen,~J.; Truhlar,~D.~G.;
  Gagliardi,~L. {Multiconfiguration Pair-Density Functional Theory}. \emph{J.
  Chem. Theory Comput.} \textbf{2014}, \emph{10}, 3669\relax
\mciteBstWouldAddEndPuncttrue
\mciteSetBstMidEndSepPunct{\mcitedefaultmidpunct}
{\mcitedefaultendpunct}{\mcitedefaultseppunct}\relax
\EndOfBibitem
\end{mcitethebibliography}
\end{document}

\listoftables

\begin{table}[tbp]
\caption{Azomethane B3LYP/6-31g(d,p) equilibrium molecular geometry in $\AA$.}
\begin{tabular}{|l|r|r|r|}
\multicolumn{4}{l}{$E=-189.283356\ E_\textrm{h}$} \\ \hline
 & \multicolumn{1}{c|}{x} & \multicolumn{1}{c|}{y} & \multicolumn{1}{c|}{z} \\ \hline
N & 0.426670 & 0.452665 & -0.000101 \\ \hline
N & -0.426702 & -0.452686 & -0.000025 \\ \hline
C & -1.797902 & 0.069203 & 0.000034 \\ \hline
H & -2.311797 & -0.329712 & -0.881217 \\ \hline
H & -2.311504 & -0.329184 & 0.881683 \\ \hline
H & -1.829142 & 1.163838 & -0.000321 \\ \hline
C & 1.797936 & -0.069186 & 0.000070 \\ \hline
H & 2.312027 & 0.330017 & -0.880973 \\ \hline
H & 2.311332 & 0.328897 & 0.881936 \\ \hline
H & 1.829109 & -1.163811 & -0.000846 \\ \hline
\end{tabular}
\label{me2n2_geom}
\end{table}

\begin{table}[tbp]
\caption{C$_2$H$_6$N$_4$ B3LYP/6-31g(d,p) equilibrium molecular geometry in $\AA$.}
\begin{tabular}{|l|r|r|r|}
\multicolumn{4}{l}{$E=-298.713635\ E_\textrm{h}$} \\ \hline
 & \multicolumn{1}{c|}{x} & \multicolumn{1}{c|}{y} & \multicolumn{1}{c|}{z} \\ \hline
H & -0.145525 & 2.534624 & 1.196098 \\ \hline
N & 0.586282 & 2.685058 & 0.454251 \\ \hline
N & 0.586282 & 1.768962 & -0.376701 \\ \hline
C & -0.376894 & 0.666619 & -0.222182 \\ \hline
H & -0.986448 & 0.788749 & 0.689250 \\ \hline
H & -1.039496 & 0.688384 & -1.095107 \\ \hline
C & 0.376894 & -0.666619 & -0.222182 \\ \hline
H & 0.986448 & -0.788749 & 0.689250 \\ \hline
H & 1.039496 & -0.688384 & -1.095107 \\ \hline
N & -0.586282 & -1.768962 & -0.376701 \\ \hline
N & -0.586282 & -2.685058 & 0.454251 \\ \hline
H & 0.145525 & -2.534624 & 1.196098 \\ \hline
\end{tabular}
\label{c2h6n4_geom}
\end{table}

\begin{table}[tbp]
\caption{C$_2$H$_4$N$_4$ B3LYP/6-31g(d,p) equilibrium molecular geometry in $\AA$.}
\begin{tabular}{|l|r|r|r|}
\multicolumn{4}{l}{$E=-297.487642\ E_\textrm{h}$}  \\ \hline
 & \multicolumn{1}{|c|}{x} & \multicolumn{1}{c|}{y} & \multicolumn{1}{c|}{z} \\ \hline
H & -2.772250 & -1.103953 & 0.000000 \\ \hline
N & -2.875835 & -0.058207 & 0.000000 \\ \hline
N & -1.761797 & 0.508543 & 0.000000 \\ \hline
C & -0.604255 & -0.293583 & 0.000000 \\ \hline
H & -0.704824 & -1.383428 & 0.000000 \\ \hline
C & 0.604255 & 0.293583 & 0.000000 \\ \hline
H & 0.704824 & 1.383428 & 0.000000 \\ \hline
N & 1.761797 & -0.508543 & 0.000000 \\ \hline
N & 2.875835 & 0.058207 & 0.000000 \\ \hline
H & 2.772250 & 1.103953 & 0.000000 \\ \hline
\end{tabular}
\label{c2h4n4_geom}
\end{table}

\begin{table}[tbp]
\caption{Azomethane electronic energy in $E_\mathrm{h}$, calculated using RHF, UHF, and CASSCF(4,4) and the 6-31g basis set at various N=N bond lengths in $\AA$.}
\begin{tabular}{|r|r|r|r|}
\hline
\multicolumn{1}{|l|}{$R_\textrm{NN}$} & \multicolumn{1}{l|}{RHF} & \multicolumn{1}{l|}{UHF} & \multicolumn{1}{l|}{CASSCF(4,4)} \\ \hline
1.0 & -187.818652 & -187.818652 & -187.872275 \\ \hline
1.1 & -187.928935 & -187.928935 & -187.993574 \\ \hline
1.2 & -187.964621 & -187.964621 & -188.041758 \\ \hline
1.3 & -187.959740 & -187.959740 & -188.050804 \\ \hline
1.4 & -187.934030 & -187.934030 & -188.040366 \\ \hline
1.5 & -187.898650 & -187.943708 & -188.021568 \\ \hline
1.6 & -187.859633 & -187.928744 & -188.000495 \\ \hline
1.7 & -187.820085 & -187.919023 & -187.980383 \\ \hline
1.8 & -187.783641 & \multicolumn{1}{l|}{ } & -187.962952 \\ \hline
1.9 & -187.762528 & \multicolumn{1}{l|}{ } & -187.949097 \\ \hline
2.0 & -187.754174 & -187.921486 & -187.939103 \\ \hline
2.1 & -187.744749 & -187.925090 & -187.932654 \\ \hline
2.2 & -187.740909 & -187.928304 & -187.928948 \\ \hline
2.3 & -187.739697 & -187.930923 & -187.927038 \\ \hline
2.4 & -187.736863 & -187.932939 & -187.926144 \\ \hline
2.5 & -187.735714 & -187.934433 & -187.925761 \\ \hline
2.6 & -187.735049 & \multicolumn{1}{l|}{ } & -187.925610 \\ \hline
2.7 & -187.734188 & -187.936255 & -187.925551 \\ \hline
2.8 & -187.733647 & -187.936767 & -187.925522 \\ \hline
2.9 & -187.733200 & -187.937106 & -187.925496 \\ \hline
3.0 & -187.732808 & \multicolumn{1}{l|}{ } & -187.925464 \\ \hline
3.1 & -187.732470 & -187.937444 & -187.925422 \\ \hline
3.2 & -187.732159 & -187.937503 & -187.925368 \\ \hline
3.3 & -187.731563 & -187.937512 & -187.925302 \\ \hline
3.4 & -187.731419 & -187.937487 & -187.925226 \\ \hline
3.5 & -187.731357 & -187.937437 & -187.925142 \\ \hline
\end{tabular}
\label{me2n2_pes1}
\end{table}

\begin{table}[tbp]
\caption{Azomethane electronic energy in $E_\mathrm{h}$, calculated using variants of DMET and LASSCF(4,4) and the 6-31g basis set at various N=N bond lengths in $\AA$.}
\begin{tabular}{|r|r|r|r|r|}
\hline
$R_\textrm{NN}$ & CAS-DMET & CAS-DMET (1e) & scCAS-DMET & LASSCF(4,4) \\ \hline
1.0 & -187.875096 & -187.871887 & -187.879578 & -187.870294 \\ \hline
1.1 & -187.997135 & -187.993043 & -188.003851 & -187.992499 \\ \hline
1.2 & -188.045907 & -188.041059 & -188.055219 & -188.041265 \\ \hline
1.3 & -188.055404 & -188.049881 & -188.067564 & -188.050552 \\ \hline
1.4 & -188.045257 & -188.039135 & -188.060503 & -188.040103 \\ \hline
1.5 & -188.026429 & -188.019920 & -188.044958 & -188.020982 \\ \hline
1.6 & -188.004869 & -187.998285 & -188.026981 & -187.999308 \\ \hline
1.7 & -187.983794 & -187.977412 &   & -187.978417 \\ \hline
1.8 & -187.964844 & -187.958952 &   & -187.960190 \\ \hline
1.9 & -187.948237 & -187.943738 &   & -187.946860 \\ \hline
2.0 & -187.917112 & -187.932047 &   & -187.937182 \\ \hline
2.1 & -187.934127 & -187.923662 &   & -187.930885 \\ \hline
2.2 & -187.926071 & -187.917973 &   & -187.927230 \\ \hline
2.3 & -187.920860 & -187.914211 &   & -187.925316 \\ \hline
2.4 & -187.916785 & -187.911692 &   & -187.924395 \\ \hline
2.5 & -187.913245 & -187.909932 &   & -187.923975 \\ \hline
2.6 & -187.909949 & -187.908631 &   & -187.923783 \\ \hline
2.7 & -187.906802 & -187.907615 &   & -187.923682 \\ \hline
2.8 & -187.903739 & -187.906789 &   & -187.923610 \\ \hline
2.9 & -187.900748 & -187.906096 &   & -187.923541 \\ \hline
3.0 & -187.897818 & -187.905502 &   & -187.923468 \\ \hline
3.1 & -187.894924 & -187.904983 &   & -187.923385 \\ \hline
3.2 & -187.892088 & -187.904524 &   & -187.923293 \\ \hline
3.3 & -187.889273 & -187.904111 &   & -187.923193 \\ \hline
3.4 & -187.886448 & -187.903738 &   & -187.923086 \\ \hline
3.5 & -187.883720 & -187.903397 &   & -187.922975 \\ \hline
\end{tabular}
\label{me2n2_pes2}
\end{table}

\begin{table}[tbp]
\caption{C$_2$H$_6$N$_4$ electronic energy in $E_\mathrm{h}$, calculated with the 6-31g basis set, calculated with various methods and the 6-31g basis set at various N=N bond lengths in $\AA$.}
\begin{tabular}{|r|r|l|r|r|r|}
\hline
$R_\textrm{NN}$ & RHF & \multicolumn{1}{r|}{UHF} & CASSCF(8,8) & CAS-DMET & LASSCF(8,8) \\ \hline
0.94 & -296.160553 & \multicolumn{1}{r|}{-296.160553} & -296.255073 & -296.263646 & -296.248733 \\ \hline
1.04 & -296.529317 & \multicolumn{1}{r|}{-296.529317} & -296.644073 & -296.657263 & -296.643922 \\ \hline
1.14 & -296.682266 & \multicolumn{1}{r|}{-296.682266} & -296.820276 & -296.840034 & -296.820078 \\ \hline
1.24 & -296.715268 & \multicolumn{1}{r|}{-296.719229} & -296.879579 & -296.908264 & -296.879306 \\ \hline
1.34 & -296.684270 & \multicolumn{1}{r|}{-296.712195} & -296.877791 & -296.918403 & -296.877509 \\ \hline
1.44 & -296.621626 & \multicolumn{1}{r|}{-296.686872} & -296.847102 & -296.903699 & -296.846649 \\ \hline
1.54 & -296.545464 & \multicolumn{1}{r|}{-296.600010} & -296.805552 & -296.884297 & -296.804856 \\ \hline
1.64 & -296.465607 & \multicolumn{1}{r|}{-296.625090} & -296.762823 & -296.878816 & -296.761870 \\ \hline
1.74 & -296.389217 & \multicolumn{1}{r|}{-296.607705} & -296.723935 & -296.941819 & -296.722753 \\ \hline
1.84 & -296.332449 & \multicolumn{1}{r|}{-296.604214} & -296.691474 & -297.028113 & -296.690120 \\ \hline
1.94 & -296.299368 & \multicolumn{1}{r|}{-296.608690} & -296.666650 & -297.031071 & -296.665195 \\ \hline
2.04 & -296.280269 & \multicolumn{1}{r|}{-296.614695} & -296.649552 & -296.996139 & -296.648091 \\ \hline
2.14 & -296.268775 & \multicolumn{1}{r|}{-296.622729} & -296.639182 & -296.949089 & -296.637617 \\ \hline
2.24 & -296.261538 & \multicolumn{1}{r|}{-296.630006} & -296.633801 & -296.903007 & -296.632159 \\ \hline
2.34 & -296.256803 & \multicolumn{1}{r|}{-296.637436} & -296.631582 & -296.861932 & -296.629891 \\ \hline
2.44 & -296.253615 & \multicolumn{1}{r|}{-296.642328} & -296.631109 & -296.824672 & -296.629389 \\ \hline
2.54 & -296.251425 &  & -296.631486 & -296.791610 & -296.629751 \\ \hline
2.64 & -296.249897 &  & -296.632218 & -296.762880 & -296.630483 \\ \hline
2.74 & -296.248820 &  & -296.633057 & -296.738109 & -296.631368 \\ \hline
2.84 & -296.248053 &  & -296.633890 & -296.716703 & -296.632178 \\ \hline
2.94 & -296.247499 &  & -296.634671 & -296.698175 & -296.632928 \\ \hline
3.04 & -296.247092 &  & -296.635387 & -296.681985 & -296.633604 \\ \hline
3.14 & -296.246783 &  & -296.636032 & -296.667920 & -296.632168 \\ \hline
3.24 & -296.246540 &  & -296.636600 & -296.656036 & -296.632712 \\ \hline
3.34 & -296.246342 &  & -296.637090 & -296.646152 & -296.632491 \\ \hline
3.44 & -296.246175 &  & -296.637505 & -296.638186 & -296.632893 \\ \hline
3.54 & -296.246035 &  & -296.637855 & -296.632218 & -296.629952 \\ \hline
3.64 & -296.245918 &  & -296.638149 & -296.627780 & -296.634377 \\ \hline
3.74 & -296.245823 &  & -296.638395 & -296.624726 & -296.627817 \\ \hline
3.84 & -296.245750 &  & -296.638604 & -296.622740 & -296.634648 \\ \hline
3.94 & -296.245699 &  & -296.638784 & -296.621730 & -296.634771 \\ \hline
4.04 & -296.245668 &  & -296.638942 & -296.621147 & -296.634879 \\ \hline
4.14 & -296.245655 &  & -296.639085 & -296.621120 & -296.635080 \\ \hline
4.24 & -296.245658 &  & -296.639217 & -296.621392 & -296.635353 \\ \hline
\end{tabular}
\label{c2h6n4_pes}
\end{table}

\begin{table}[tbp]
\caption{C$_2$H$_4$N$_4$ electronic energy in $E_\mathrm{h}$, calculated with the 6-31g basis set, calculated with various methods and the 6-31g basis set at various N=N bond lengths in $\AA$.}
\begin{tabular}{|r|r|r|r|r|}
\hline
$R\_\textrm{NN}$ & \multicolumn{1}{l|}{CASSCF(8,8)} & \multicolumn{1}{l|}{CASSCF(10,10)} & \multicolumn{1}{l|}{LASSCF(8,8)} & \multicolumn{1}{l|}{LASSCF(10,10)} \\ \hline
0.75 & -293.076181 & -293.108096 & -293.071040 & -293.065354 \\ \hline
0.85 & -294.406532 & -294.443629 & -294.400658 & -294.412511 \\ \hline
0.95 & -295.106005 & -295.148336 & -295.114845 & -295.136153 \\ \hline
1.05 & -295.448328 & -295.515061 & -295.475085 & -295.501768 \\ \hline
1.15 & -295.643904 & -295.676967 & -295.636323 & -295.665612 \\ \hline
1.25 & -295.696348 & -295.730227 & -295.688324 & -295.718426 \\ \hline
1.35 & -295.692075 & -295.727245 & -295.683394 & -295.713125 \\ \hline
1.45 & -295.661229 & -295.698325 & -295.651731 & -295.680276 \\ \hline
1.55 & -295.620718 & -295.660617 & -295.610249 & -295.636998 \\ \hline
1.65 & -295.579720 & -295.623626 & -295.568034 & -295.592479 \\ \hline
1.75 & -295.543404 & -295.592525 & -295.529823 & -295.551479 \\ \hline
1.85 & -295.516144 & -295.569256 & -295.498063 & -295.516452 \\ \hline
1.95 & -295.498781 & -295.552968 & -295.473876 & -295.490725 \\ \hline
2.05 & -295.487615 & -295.542016 & -295.457296 & -295.474328 \\ \hline
2.15 & -295.480800 & -295.535171 & -295.447345 & -295.464499 \\ \hline
2.25 & -295.477157 & -295.531409 & -295.442214 & -295.459435 \\ \hline
2.35 & -295.475649 & -295.529750 & -295.440103 & -295.457356 \\ \hline
2.45 & -295.475421 & -295.529361 & -295.439644 & -295.456915 \\ \hline
2.55 & -295.475881 & -295.529653 & -295.439980 & -295.457264 \\ \hline
2.65 & -295.476682 & -295.530274 & -295.440642 & -295.457936 \\ \hline
2.75 & -295.477654 & -295.531042 & -295.441395 & -295.458699 \\ \hline
2.85 & -295.478714 & -295.531876 & -295.442131 & -295.459446 \\ \hline
2.95 & -295.479789 & -295.532741 & -295.442805 & -295.460130 \\ \hline
3.05 & -295.480798 & -295.533596 & -295.443402 & -295.460737 \\ \hline
3.15 & -295.481680 & -295.534387 & -295.443918 & -295.461262 \\ \hline
3.25 & -295.482414 & -295.535074 & -295.444355 & -295.461708 \\ \hline
3.35 & -295.483007 & -295.535644 & -295.444719 & -295.462080 \\ \hline
3.45 & -295.483479 & -295.536104 & -295.445016 & -295.462386 \\ \hline
3.55 & -295.483852 & -295.536471 & -295.445253 & -295.462635 \\ \hline
3.65 & -295.484149 & -295.536764 & -295.445449 & -295.462838 \\ \hline
3.75 & -295.484389 & -295.537001 & -295.445701 & -295.463005 \\ \hline
3.85 & -295.484587 & -295.537197 & -295.445707 & -295.463146 \\ \hline
3.95 & -295.484754 & -295.537364 & -295.445870 & -295.463267 \\ \hline
4.05 & -295.484901 & -295.537510 & -295.445944 & -295.463377 \\ \hline
4.15 & -295.485034 & -295.537642 & -295.445996 & -295.463478 \\ \hline
4.25 & -295.485156 & -295.537763 & -295.446321 & -295.463571 \\ \hline
\end{tabular}
\label{c2h4n4_pes}
\end{table}